\def\gtap{\mathrel{\hbox{\rlap{\lower.55ex \hbox {$\sim$}}
                   \kern-.3em \raise.4ex \hbox{$>$}}}}
\def\ltap{\mathrel{\hbox{\rlap{\lower.55ex \hbox {$\sim$}}
                 \kern-.3em \raise.4ex \hbox{$<$}}}}

\newcommand{\lsun}{L_{\odot}}

\newcommand{\ergsec}{\mbox{erg s$^{-1}$}}

\newcommand{\chandra}{Chandra}

\newcommand{\Lx}{\mbox{$L_{\rm X}$}}
\newcommand{\Msun}{\mbox{$M_{\odot}$}}
\newcommand{\rsun}{\mbox{$R_{\odot}$}}

\newcommand{\gc}{globular cluster}
\newcommand{\gcs}{globular clusters}
\newcommand{\lmxbs}{LMXBs}
\newcommand{\lum}{luminosity}
\newcommand{\lums}{luminosities}
\newcommand{\bh}{black hole}

\newcommand{\gal}{galaxy}
\newcommand{\gals}{galaxies}

\newcommand{\elgals}{elliptical galaxies}

\newcommand{\SN}{specific frequency}

\newcommand{\lmxb}{LMXB}
\newcommand{\lmns}{LMXB$^{\mathrm NS}$}
\newcommand{\lmbh}{LMXB$^{\mathrm BH}$}

\documentclass{caps}
\usepackage{psfig,graphics,amssymb}

\def\fv{}
\def\mk{}

\begin{document}

\pagenumbering{roman}
\cleardoublepage

\pagenumbering{arabic}

\setcounter{chapter}{7}

\setcounter{table}{0}

\author[Frank Verbunt \&\ Walter H.G.\ Lewin]{Frank Verbunt\\
Astronomical Institute, Postbox 80.000, 3508 TA Utrecht, The Netherlands\\
\and
Walter H.G.\ Lewin\\Massachusetts Institute of Technology, Physics Department, Center for Space Research\\ Cambridge, MA 02139, USA
}

\chapter{Globular Cluster X-ray Sources}

\section{Introduction: Some historical remarks}

The earliest detections of luminous X-ray sources ($\Lx \gtap
10^{36}$~\ergsec) in globular clusters were made with the Uhuru
and OSO-7 Observatories (Giacconi et al.\ 1972 \& 1974; Clark,
Markert \& Li, 1975; Canizares \& Neighbours, 1975). About 10\%\ of the
luminous X-ray sources in our Galaxy are found in globular clusters.
This implies that the probability (per
unit mass) of finding a luminous X-ray source in a globular cluster is
about two to three orders of magnitude higher than of finding one in the
rest of our galaxy (Gursky 1973, Katz 1975). 
Clearly, the conditions in globular clusters are
very special in that they must be very efficient breeding grounds for
X-ray binaries.  For reviews which reflect the ideas in
the late seventies and early eighties, see Lewin (1980), Lewin \& Joss
(1983), Van den Heuvel (1983) and Verbunt \&\ Hut (1987). 
At that time there was no evidence
for a substantial population of binaries in globular clusters;
e.g.\ Gunn \& Griffin (1979) did not find a single binary in a spectroscopic
search for radial velocity variations of 111 bright stars in M\,3.

Clark (1975) suggested that the luminous cluster sources are binaries
formed by capture from the remnants of massive stars.  Fabian, Pringle
\& Rees (1975) specified that they are formed via tidal capture of
neutron stars in close encounters with main-sequence stars. 
Sutantyo (1975) suggested direct collisions between giants and
neutron stars as a formation mechanism. Hills (1976) examined the
formation of binary systems through star-exchange interactions between
neutron stars and primordial binaries of low-mass stars. Hut \& Verbunt (1983)
compared the relative efficiencies of tidal capture and exchange
encounters for neutron stars and for white dwarfs; and showed that
the distribution of X-ray sources among globular clusters with
different central densities and core sizes is compatible with
the formation by close encounters (Verbunt \&\ Hut 1987). The importance of
mass segregation, which drives the neutron stars to the core,
thereby enhancing the capture rate, was demonstrated 
by Verbunt \&\ Meylan (1988).

As can be seen from the discovery references in Table 8.1,
five luminous globular cluster X-ray sources were known by 1975,
eight by 1980, ten by 1982, and thirteen to date.
Twelve of these have shown type~I X-ray bursts. Measurements of the
black-body radii of the burst sources indicated that they are neutron
stars (Swank et al.\ 1977; Hoffman, Lewin \& Doty 1977a \& 1977b; Van
Paradijs 1978). For a review, see Lewin, Van Paradijs \& Taam
(1995); see also \S3\mk. The absence of luminous accreting black holes in clusters of
our galaxy is presumably a consequence of the small total number of
sources, as discussed in \S\ref{vlsecbh}. There is \mk
growing evidence that black-hole binaries may exist in globular
clusters in several elliptical galaxies (see \S8.3).

Because of the observed correlation between the occurrence of a luminous
X-ray source in a globular cluster and a \mk high central density, it
was expected already early on that these luminous sources would be
located close to the cluster centers. These expectations were
confirmed by measurements, carried out with the SAS-3 X-ray
Observatory, which showed that the positional error circles with radii
of 20--30 arcsec (90\% confidence) included the optical centers of the
clusters (Jernigan \& Clark 1979). Later work with the Einstein
observatory greatly refined the positional measurements (Grindlay et
al.\ 1984). Bahcall \& Wolf (1976) have shown that under certain 
assumptions, the average mass of the X-ray sources can be derived from
their positions with respect to the cluster center.
Even if one accepts the assumptions made,
the average mass derived this way for the luminous X-ray
sources in \gcs\ was not sufficiently accurate to classify these
sources, but the result
was consistent with the earlier conclusions (see e.g., Lewin 1980; Lewin \& Joss 1983) that these are accreting
neutron stars (Grindlay et al.\ 1984).

\begin{table*}
\centerline{
\begin{tabular}{l@{ }ll@{ }lllll}
cluster   & position  & discovery         & 1st burst    &
 $M_{\lambda}$   &   $P_b$ & TOXB \\   
NGC\,$1851$ & 0512$-$40\,\cite{ham+01} & O7\cite{cml75} & UH\cite{fj76} &
 5.6B\cite{damd96} & & \phantom{T}UUU\\
NGC\,$6440$ & 1745$-$20\,\cite{plv+02} & O7\cite{mbc+75} & BS\cite{zvs+99} &
 3.7B\cite{vkzh00} & & T\phantom{}$-$N$-$& \\
NGC\,$6441$ & 1746$-$37\,\cite{ham+02} & UH\cite{gmg+74} & EX\cite{sfp+87} &
 2.4B\cite{damd98} & 5.7hr\cite{sdal93} &  \phantom{T}$-$NN \\
NGC\,$6624$ & 1820$-$30\,\cite{ksa+93} & UH\cite{gmg+74} & ANS\cite{ggs+76} &
 3.0B\cite{amd+97} & 11.4m\cite{spw87} & \phantom{T}UUU \\
NGC\,$6652$ & 1836$-$33\,\cite{heg01} & H2\cite{hw85} & BS\cite{zvh+98} &
 5.6B\cite{heg01} & c & \phantom{T}UUU\\
NGC\,$6712$ & 1850$-$09\,\cite{ghs+84} & AV\cite{sptp76} & S3\cite{hcl80} &
 4.5B\cite{hcn+96} &  20.6m$^d$\cite{hcn+96} & \phantom{T}UUU \\
NGC\,$7078$-1$^a$ & 2127+12\,\cite{wa01} & Ch\cite{wa01} & & 
 0.7B\cite{aft84} & 17.1hr\cite{iak+93} & \phantom{T}$-$$-$$-$\\
NGC\,$7078$-2$^a$ & 2127+12\,\cite{wa01} & Ch\cite{wa01} & Gi\cite{dim+90} &
 3.1U\cite{wa01} & & \phantom{T}$-$$-$U \\
Terzan\,1  & 1732$-$30\,\cite{jvh95} & Ha\cite{moi+81} & Ha\cite{moi+81} 
      &               & & T\phantom{}$-$$-$$-$   \\
Terzan\,2   & 1724$-$31\,\cite{ghs+84} & O8\cite{sbb+77} & O8\cite{sbb+77} &
& & \phantom{T}$-$NU\\
Terzan\,5  & 1745$-$25\,\cite{heg+03} & Ha\cite{moi+81} & Ha\cite{moi+81} &
1.7J\cite{heg+03} & & T\phantom{}$-$U$-$ \\
Terzan\,6  & 1751$-$31\,\cite{zhm+03} & RO\cite{phv91} & BS\cite{zhm+03}&
                & 12.36h\cite{zbc+00} & T\phantom{}$-$N$-$ \\
Liller\,1  & 1730$-$33\,\cite{hdam01} & S3$^b$\cite{ldc+76} & S3\cite{hml78} 
& & & T\phantom{}$-$$-$$-$ \\
\end{tabular}
}
$^a$A luminous X-ray source in NGC\,7078 was already found with Uhuru\cite{gmg+74},
the Chandra observations resolved this source into two sources\\
$^b$X-ray source (the Rapid Burster) discovered before the globular cluster!\\
$^c$43.6\,m period originally assigned to this source is period of 
fainter X-ray source \cite{heg01}\\
$^d$or the alias period of 13.2\,m

\caption[o]{\it Some information on the luminous X-ray sources in
globular clusters of our galaxy. Columns from left to right (1)
cluster, (2) rough position (B1950, often used as source name) with
reference to the currently most accurate position, (3) the satellite
with which the source was discovered as a cluster source, (4) the
satellite which detected the first type~I X-ray burst from the source, (5) absolute
magnitude with filter of optical counterpart, (6) orbital period, (7)
indication (with a ``T'') whether the source is a transient. The last
three columns indicate whether a normal (N) or ultrashort (U) orbital period
is suggested by the comparison of optical with X-ray luminosity (column 8,
under ``O''), the X-ray spectrum (column 9, under ``X'') and the maximum flux
reached during bursts (column 10, under ``B''). A ``$-$'' in columns 8-10 indicates
that no information is available. Satellite names are abbreviated as
O(SO-)7, O(SO-)8, UH(URU), H(EAO-)2, A(riel-)V, Ch(andra), Ha(kucho),
RO(SAT), S(AS-)3, B(eppo)S(ax), EX(OSAT), Gi(nga). Note that the
absolute magnitudes are subject to uncertainties in distance and
reddening; also most sources are variable (see Deutsch et al.\
2000). \nocite{dma00}
\label{tbright}
}
\end{table*}

\begin{figure}
\centerline{
\parbox[b]{7.0cm}{\psfig{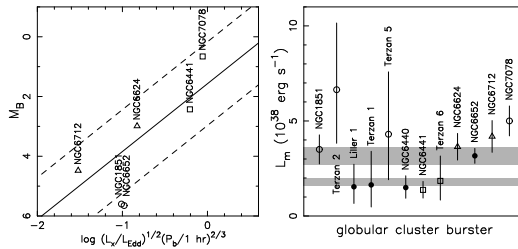}}
\parbox[b]{5.0cm}{\caption[o]{\it Various types of X-ray sources in
globular clusters; sg, ms, wd, and ns stand for subgiant,
main-sequence star, white dwarf, and neutron star, respectively. From
top to bottom: luminous low-mass X-ray binary, low-luminosity low-mass
X-ray binary, recycled radio pulsar (here with a white dwarf
companion), cataclysmic variable, and magnetically active
binary. $L_{sd}$ stands for spin-down luminosity. Approximate maximum
luminosities (in the 0.5-4.5 keV range) are indicated on the right.
The low-mass X-ray binaries harboring a
neutron star are referred to as \lmns; when they harbor a back
hole, we refer to them as \lmbh, and we refer to both groups together
as \lmxb. \label{verbfb}}
}
}
\end{figure}

Sources with $L_x\ltap10^{35}$\ergsec\ were first found in globular clusters
with Einstein (Hertz \&\ Grindlay 1983). More were found with ROSAT,
by a variety of authors (Table\,\ref{tdim}); a final,
homogeneous analysis of the complete ROSAT data was made by Verbunt (2001).
On the basis of these Einstein and ROSAT results, it has gradually
become clear that these sources are a mix of various types (see Figure\,\ref{verbfb}).
Hertz \&\ Grindlay (1983) suggested that they were mainly
cataclysmic variables, and noted that the low-luminosity source in NGC\,6440
could be the quiescent counterpart of the luminous transient source in
that cluster. Verbunt et al.\ (1984) argued that the more luminous
of the low-luminosity sources are all quiescent low-mass X-ray binaries.
The first radio \mk pulsar detected as a low-luminosity source in a globular cluster is
the pulsar in NGC\,6626 (M\,28, Saito et al.\ 1997). Finally, Bailyn et al.\
(1990) pointed out that magnetically active binaries also reach 
X-ray luminosities in the range of the less luminous sources
detected with ROSAT.

It was also realized that some of the sources could be unresolved
multiple sources; and unresolved emission was found e.g.\ by Fox 
et al.\ (1996) in NGC\,6341 and NGC\,6205.
However, it is fair to say that the actual plethora of sources
shown by the Chandra observations in virtually every cluster
that it observed (Tables\,\ref{tdim}, \ref{verbta}) was unpredicted.
These observations confirmed that quiescent low-mass X-ray
binaries, cataclysmic variables, pulsars, and magnetically active
binaries are all X-ray sources in globular clusters, as is
discussed in \S\ref{vlsecfs}. 
Whereas some of the Einstein and ROSAT sources
are confirmed with Chandra as single sources, others have been
resolved into multiple sources; details are given in Table 8.2.

The positions obtained with ROSAT were sufficiently accurate to
find plausible optical counterparts in Hubble Space Telescope (HST) \mk observations in a number
of cases. This work was pioneered in NGC\,6397 with a search for
$H\alpha$ emitting objects by Cool et al.\ (1993, 1995), and spectroscopic
follow-up \mk confirming the classification as cataclysmic variables
by Grindlay et al.\ (1995), Cool et al.\ (1998) and Edmonds et
al.\ (1999). Plausible candidate counterparts were also found for
two X-ray sources in the core of $\omega$\,Cen (Carson et al.\ 2000).
All of these suggested counterparts were confirmed with the more
accurate positions obtained with Chandra. In 47\,Tuc, of the 
candidate counterparts suggested by a variety of authors,
Verbunt \&\ Hasinger (1998) only retain three, on the basis
of more accurate positions of the X-ray sources; these were also
confirmed with Chandra. Ferraro et al.\ (1997) suggested 
ultraviolet stars as counterparts for two sources found by Fox et 
al.\ (1996) in NGC\,6205 (see also Verbunt 2001). An ultraviolet
counterpart suggested by Ferraro et al.\ (2000) for a source in 
NGC\,6341 is incompatible with the position of that source
(Geffert 1998, Verbunt 2001).
Another approach is to look for X-rays from an already known
special object. A dwarf nova known since 1941 well outside the central 
region of NGC\,5904 (Oosterhoff 1941) was detected with ROSAT 
(Hakala et al.\ 1997), and a pulsar in M28 (Lyne et al.\ 1987) was
detected with ASCA (Saito et al.\ 1997). Before Chandra, no
magnetically active binary was suggested as an optical counterpart 
for a specific X-ray source.

\begin{table*}
\centerline{
\begin{tabular}{lllll}
cluster   & E & R & C/X & comments \\
NGC\,104/47Tuc & 1 &   5+4 & 39+66 & E=R9=C42 (CV) R5=C58 R7=C46\\
 & & & &                  R6=C56 R10=C27 R11=C25 R13=C2\\
 & & & &                  R19=C30 R4 outside C-frame\\
NGC\,288       & & 1 & & \cite{sir+99} \\
NGC\,362 &       & 2  \\
Pal 2   &     & 0+1 & & \cite{rdlm94} \\
NGC\,1904/M79 & 1 & 0+1 & &  E=R \\
NGC\,5139/$\omega$Cen  &  1+4 & 3+3 & 3+97 & \cite{vj00} core: EC$>$(R9a=C6/R9b=C4) \\
 & & &                  & both CVs; out-of-core: EB=R7=C3 qLMXB; \\
 & & &                  & EA=R3, ED=R4, EE=R5 foreground stars\\
NGC\,5272/M3 & 1 &   1  & & \cite{dag99} E=R, CV/SSS? also \cite{hgb93}, opt.id.\ \cite{ekh04} \\
NGC\,5824    & 1 &   0 & & R limit just below E detection level \\
NGC\,5904/M5    & & 0+1 & 10$^a$ &  \cite{hcjv97} \\
NGC\,6093/M80 & & 1  & 9+10 & R$>$(C1/C2/C4/C7..) \\
NGC\,6121/M4    & & 1 & 12+19 & R=C1 \\
NGC\,6139      &  & 1  \\
NGC\,6205/M13  & & 2+1 & 2+1 & core: RGa=X3 qLMXB; RGb$\neq$X \\
 & & & &                X2$\neq$R out-of-core: RF=X6 \\
NGC\,6266/M62 & &     1 & 45$^a$  \\
NGC\,6341/M92 & &     1  & & \cite{jvh94}, \cite{flm+96} \\
NGC\,6352     & &     0+1 & & \cite{jvh96} \\
NGC\,6366     & &     1 & 1$^a$ & \cite{jvh96}  \\
NGC\,6388   & & 0+1  \\
NGC\,6397   & & 5+1 & 9+11 & R4a/b/c/d/e=C19/17/23/22/18 R13=C24\\
NGC\,6440 & 1 & 2 & 24 & E$>$(R1$>$C2/C4/C5..,R2$>$C1/C3...) \\ 
NGC\,6541  & 1 & 1   \\
NGC\,6626/M28 & & 3+1  & 12+34 & core: (R2a+2b)=C26 R2c=C19 \\
 & & & &               out-of-core: R7=C17\\
NGC\,6656/M22 & 1+3 & 1 & 3+24 & core: E=R=X16, opt id.\cite{ack03} X18$\neq$R \\
 & & & &             out-of-core: E prob.\ not related to cluster\\
NGC\,6752 & & 4+2 & 9+8 & core: R7a$>$C4/7/9 R21$>$C11/12/18 R7b=C1 \\
 & & &  &             R22=C6 out-of-core: R6=C3 R14=C2\\
NGC\,6809 & & 1  & & \cite{jvh96} \\
NGC\,7099 & & 0+1  & 5$^a$ & \cite{jvh94} \\
\\
total:  &    8+7 & 37+18 \\
\end{tabular}
}
$^a$number within half-mass radius from \cite{pla+03}, 
detailed analysis not yet published

\caption[o]{\it Observations of low-luminosity sources in globular clusters.
We list the 
number of sources found with Einstein (under E), ROSAT (R) and Chandra or
XMM (C/X). Numbers following the + sign indicate sources outside the
cluster core. Note that the detection limits are very different between 
clusters. 
References are Hertz \&\ Grindlay (1983) for Einstein sources,
Verbunt (2001) and references therein for ROSAT sources. 
References for Chandra and XMM-Newton are listed in Table\,\ref{verbta}. 
Under comments we
provide additional references for ROSAT, give occasional 
source types quiescent (i.e.\ low-luminosity)
low-mass X-ray binary (qLMXB) and cataclysmic variable (CV), and
indicate the relation between sources observed by subsequent satellites.
$=$ identical; $>$ resolved into multiple sources; A$\neq$B source A
not detected by satellite B, due to significant variability.
Source numbers under comments are those in the references given.
 \label{tdim}
}
\end{table*}

The luminous \mbox{X-ray} sources in globular clusters are binary
systems, and most (if not all) of the low-luminosity 
X-ray sources are also binary systems
or have evolved from them.  The presence of binaries is a very
important factor in the evolution of a globular cluster (Hut et al.\
1992). Theoretical considerations and numerical calculations show that
a cluster of single stars is unstable against collapse of its core
(H\'enon 1961). If binaries are present, however, close binary-single
star encounters can increase the velocity of the single stars by
shrinking the binary orbits.  Binaries can therefore become a
substantial source of energy for the cluster, sufficient even to
reverse the core collapse.  Even a handful of very close binaries can
significantly modify the evolution of a globular cluster (Goodman \&
Hut, 1989).  With a million stars in the cluster as a whole, the
number of stars in the core of a collapsed cluster may be only a few
thousand.  A close binary system, such as an \mbox{X-ray} binary, will
have a binding energy that can easily be a few hundred times larger
than the kinetic energy of a single star.  A dozen such systems, as
they were formed, released an amount of energy that is comparable to
the kinetic energy of the core as a whole.  Encounters between such
binaries and other single stars or binaries have the potential to
change the state of the core dramatically by increasing or decreasing
the core size, and by kicking stars and binaries into the cluster halo
or even out of the cluster altogether. The study of the binaries, and
\mbox{X-ray} binaries in particular, is therefore of great importance
as they play a key role in the cluster's dynamical evolution.

It has been suggested that globular clusters are responsible for the
formation of all or some of the low-mass X-ray binaries in our Galaxy,
also those outside clusters now (e.g.\ Grindlay \&\ Hertz 1985).
Specifically, such an
origin was suggested by Mirabel et al.\ (2001) for the black-hole
X-ray binary XTE\,J1118+480, and by Mirabel \&\ Rodrigues (2003)
for Sco X-1, on the basis of their orbits in the Galaxy.
The discovery of very large populations of cluster
X-ray sources in other galaxies has rekindled the question of
cluster origin for non-cluster sources. In 
\S\ref{secfield}
we will give the \mk reasons why we believe that most X-ray binaries
in the disk of our Galaxy were formed there, and not
in globular clusters. 
\nocite{gh85}\nocite{mdm+01}\nocite{mr03}

\section{The luminous \gc\ X-ray sources in the Galaxy}
In Table 8.1, we list some information on the 13 luminous \gc\ X-ray
binaries in the Galaxy. A comprehensive study of the X-ray spectra of
these luminous sources is made by Sidoli et al.\ (2001), who used
BeppoSAX observations in the spectral range \mk between 0.1 and 100 keV.  They
find that the luminous sources in NGC\,1851, NGC\,6712 and NGC\,6624
have similar spectra. When a two-component model (the sum of a
disk-blackbody and a Comptonized spectrum) is used to describe the
spectrum, the fitted radii and temperatures are compatible with values
expected for radii and temperatures of the inner disk. The spectrum of
the luminous source in NGC\,6652 is similar, except that some radiation
is blocked, possibly by the outer disk (Parmar et al.\ 2001). The spectra of the luminous sources in NGC\,6440, NGC\,6441,
Terzan\,2 and Terzan\,6 are very different. In the two-component
model the inner disk temperature was higher than that of the seed
spectrum injected into the Comptonizing plasma, and the inner radius
was smaller than those of realistic neutron-star radii. BeppoSAX
observed the Rapid Burster in Liller\,1 and the luminous source in
Terzan 1 when these sources were in a low state; the two luminous sources
in NGC\,7078 could not be resolved.

Sidoli et al.\ (2001) suggest, on the basis of binary systems whose
orbital periods are known (see Table 8.1), that the two types of
spectra correspond to two types of orbital periods: the
ultrashort-period systems (observed in NGC\,6712 and NGC\,6624) and
the longer/normal period systems (observed in NGC\,6441 and
NGC\,7078-1). We classify the sources as ultrashort 
(orbital period less than 60\,m, say)\fv\ or normal 
based on
this correspondence, in Table\,\ref{tbright}, column (9). It may be
noted that this classification does not depend on the physical
interpretation of the spectra. The luminous source in Terzan 5 has
been added to the suggested ultrashort-period systems, on the basis of
its X-ray spectrum as observed with Chandra (Heinke et al.\ 2003a).
\nocite{spo+01}\nocite{pos+01}\nocite{heg+03}

It is interesting to compare this tentative classification with two
others.  The first of these is based on the finding that
ultrashort-period systems have a much lower ratio of optical to X-ray
flux than systems with longer periods: the optical flux is due to
reprocessing of X-rays in the accretion disk; a small accretion
disk has therefore a small optical flux (Van Paradijs \&\ McClintock
1994). Thus the absolute visual magnitude, in conjunction with the
X-ray luminosity, may be used to estimate whether the orbital period
is ultrashort or not. This is done in column (8) of Table\,\ref{tbright}.
 The other tentative classification scheme is
based on the notion that the white dwarf donor stars in
ultrashort-period systems do not contain hydrogen.  The X-ray bursts
of hydrogen-free matter can reach higher luminosities because the
Eddington limit is higher in the absence of hydrogen. Kuulkers et al.\
(2003) have carefully investigated the maximum observed luminosities
of bursters in globular clusters. On this basis we can also
tentatively classify ultrashort-period systems, as we have done in
column(10) of Table\,\ref{tbright}.
\nocite{vpm94}\nocite{khz+03}

It is seen that the different classifications are consistent for known
ultrashort-period systems in NGC\,6624 and NGC\,6712 and tentatively
classified ultrashort systems in NGC\,1851 and NGC\,6652 and for the
systems with known longer period in NGC\,6441 and Terzan 6.  Two
tentative indicators for the source in Terzan 2 are contradictory.

Five of the thirteen luminous X-ray sources are transients. The source in
Terzan 1 has been consistently luminous until about 1999, when it switched
off (Guainazzi et al.\ 1999).  The Rapid Burster in Liller\,1 and the
luminous source in Terzan 6 are recurrent transients, showing outbursts
quite frequently. Intervals of $\sim$6-8 months (Lewin et al.\ 1995)
and $\sim$100 days (Masetti 2002) were observed for the Rapid
Burster, and $\sim$4.5 months for the luminous source in Terzan 6 (in 't
Zand et al.\ 2003).  The luminous source in NGC\,6440 is a transient
whose outbursts have been detected in 1971, 1998 and 2001 (see
\S8.2.1).  The transient source in Terzan 5 entered a rare high
state in August 2000 (Heinke et al.\ 2003a, and references
therein).   Interestingly, most (known and suggested) ultrashort-period systems
  are persistent sources. Note, however, that one of the two periods
  known for low-luminosity \fv
low-mass X-ray binaries is also ultrashort (see
  Fig.\,\ref{verbfper}). (The source in NGC\,6652 does
occasionally drop below $\sim$$10^{36}$
\ergsec, but it is not known by how much.) Whether the above
correlations are significant remains to be seen, and \mk will only
become evident once more secure orbital periods have been
determined. 

\begin{figure}
\centerline{\psfig{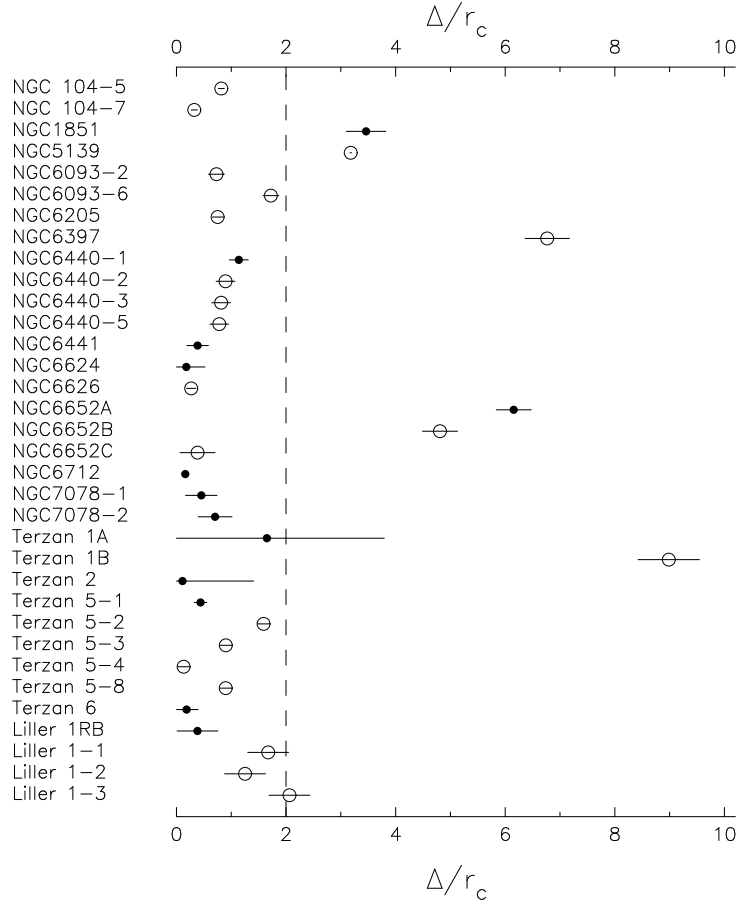}}

\caption[o]{\it Distance $\Delta$ of Low-Mass X-ray Binaries
to the center of the globular cluster in which they are located,
in units of the core radius $r_c$. Luminous and low-luminosity sources
are indicated with $\bullet$ and $\circ$, respectively. Errors
are computed from the uncertainty in the X-ray position and from the uncertainty
in the position of the cluster center (assumed to be $1.2''$).
Core radii and centers are
taken from Harris (1996, February 2003 version), except for Terzan 6
(in 't Zand et al.\ 2003). References for the X-ray positions are
in Tables\,\ref{tbright} and \ref{tdim}. It is seen that
most, but not all, X-ray binaries are within 2$r_c$. \label{verbfrrc}}

\end{figure}

With Chandra, the positions of the luminous sources have become more
accurate. In Figure\,\ref{verbfrrc} we show these positions, together
with those of the low-luminosity sources that also contain a neutron
star. It is seen that some sources, e.g.\ the luminous source in
NGC\,6652, are at a large distance from the cluster core.

\subsection{Notes on individual sources}

{\it NGC\,1851}. The accurate Chandra position for the luminous source
in
NGC 1851 confirms the previously suggested optical counterpart; this
star is very faint, considering the brightness of the X-ray source,
which suggests that the binary is an ultra-short period binary (see
\S8.2 and Table 8.1), i.e.\ $P_b<1$\,h (Homer et al.\ 2001a).
\nocite{ham+01}

{\it NGC\,6440}. The luminous source in NGC\,6440 is a transient;
outbursts were detected in 1971 with OSO-7 and Uhuru (Markert et al.\ 1975,
Forman et al.\ 1976), and again in 1998 and 2001 with BeppoSAX (in 't
Zand et al.\ 1999, 2001).  The 1998 outburst was followed up with NTT
and VLT observations. An optical transient was found at the
approximate location of the X-ray transient (Verbunt et al.\
2000). The 2001 outburst was observed with Chandra (in 't Zand et al.\
2001), and the source was identified with one of four low-luminosity
sources found earlier by Pooley et al.\ (2002b). The 1998 optical and
the 2001 X-ray transient are the same source.
\nocite{fjt76}\nocite{plv+02}\nocite{vkzh00}\nocite{zvs+99}\nocite{zkp+01}

{\it NGC\,6624}. The luminous source in NGC\,6624 has an orbital
period of 865 s, indicating that the donor is a white dwarf (Verbunt
1987). For such a donor, theory predicts that the orbital period
increases with time: $\dot P_b/P_b>8.8\times 10^{-8}{\rm yr}^{-1}$.
However, observations made in the period 1967 to 1997 show a decrease
in the period, of order $\dot P_b/P_b=-5.3\times 10^{-8}{\rm yr}^{-1}$
(Van der Klis et al.\ 1993, Chou \& Grindlay 2001). This continued
decrease cannot be explained by changes in the disk size. However, the
X-ray source is located close to the center of the cluster (King et
al.\ 1993), and if the central density is high enough, acceleration of
the binary in the cluster potential may explain the difference (Chou
\& Grindlay 2001). Further study is required as discrepancies exist
between reported positions for the cluster's center. It is 
important that the central density of the cluster be determined more
accurately.  A viable alternative may be that the donor is not a white
dwarf, but a stripped core of a slightly evolved main-sequence star
(Podsiadlowski, Rappaport \& Pfahl 2002).

{\it NGC\,6652}. Chandra observations of NGC\,6652 show three
low-luminosity sources in addition to the luminous source. The optical
counterpart with a 43.6\,m orbital period 
previously suggested (Deutsch et al.\ 2000) for the luminous
source turns out to be one of the low-luminosity sources instead (Heinke et al.\
2001). The Chandra data were obtained with the High Resolution Camera
and thus do not
contain much spectral information.  The visual brightness of the new
optical counterpart of the luminous X-ray source is still very low; and the
suggestion (Deutsch et al.\ 2000) that this source is an ultra-short period binary stands
(see Table 8.1).
\nocite{dma00}\nocite{heg01}

{\it NGC\,7078}. A Chandra observation of NGC\,7078 (M15) showed that
this cluster contains two luminous sources, at a separation of 3$''$,
seen as a single source in earlier observations with instruments that
have less spatial resolution (White \& Angelini 2001). The presence
of two sources actually had been predicted by Grindlay (1992), as a
solution to a
puzzle posed by previous observations. The high optical to X-ray
flux ratio indicated that the central X-ray source is hidden by the
accretion disk, and that only X-rays scattered in our direction by a
corona are detected; this implies that the intrinsic X-ray luminosity
exceeds the observed luminosity by almost two orders of magnitude
(Auri\`ere et al.\ 1984). However, burst observations indicated that
the bursts reached the Eddington limit for the distance to M15; this
implied that there was no blockage of radiation, and thus that the
observed persistent flux was representative for the full luminosity
(Dotani et al.\ 1990). The brightest of the two (7078-2, see Table 8.1) is
the burster; the optical counterpart is probably a blue star with
$U=18.6$; its position is determined most accurately from its radio
counterpart (Kulkarni et al.\ 1990).  The less luminous source 7078-1
has the disk corona, and is identified optically with a 17.1 hr
partially eclipsing binary (Ilovaisky et al.\ 1993). Its optical
brightness and the orbital period -- revealed by variable, non-total
eclipses -- indicate that the donor in this system is a
sub-giant. Ultraviolet lines with strong P~Cygni profiles indicate
extensive mass loss. An analysis of the eclipse timing
puts a rough upper limit on the period change of 0.01~d in 22~yr
(Naylor et al.\ 1992; Ioannou et al.\ 2003).  An extreme ultraviolet
flux has been detected from M15. It was believed to come from the
X-ray binary AC211, the optical counterpart of 7078-1 (Callanan et
al.\ 1999). We suggest that some UV may also come from 7078-2 which
allows for a direct view to the center of the accretion disk.
\nocite{wa01}\nocite{aft84}\nocite{dim+90}\nocite{cdf99}\nocite{kgwm90}
\nocite{nch+92}\nocite{iak+93}\nocite{izn+03}

{\it Terzan 1}. When Terzan 1 was observed with BeppoSAX in April 1999, the
luminosity had dropped to about $2\times10^{33}$ \ergsec, indicating
that the luminous source in this cluster had gone into quiescence
(Guainazzi et al.\ 1999).  Accurate positions for the luminous source
had been obtained with EXOSAT ($8''$ accuracy, Parmar et al.\ 1989)
and ROSAT ($5''$ accuracy, Johnston et al.\ 1995); remarkably, the
source detected with Chandra is not compatible with these positions
(Wijnands et al.\ 2002).  Probably, all observations of the bright
state before 1995 refer to the same source, since the detected
luminosities are all similar at, or just below, $10^{36}$\ergsec\
(Skinner et al.\ 1987; Parmar et al.\ 1989, Verbunt et al.\ 1995,
Johnston et al.\ 1995). This source was discovered in 1980 during
observations with Hakucho; only two bursts were observed in one
week. The upper limit to the persistent flux was
$\sim$$10^{36}$\ergsec\ (Makishima et al.\ 1981). 
It is not clear whether BeppoSAX detected
the faint state of the luminous source, or the low-luminosity
source found with Chandra.
\nocite{gpo99}\nocite{psg89}\nocite{jvh95}\nocite{whg02}\nocite{swe+87}
\nocite{vbhj95}\nocite{moi+81}

{\it Terzan 5}. Observations of Terzan 5 with \chandra\ show nine sources in
addition to the transient; four of these are probably low-luminosity \lmns\
 (Heinke et al.\ 2003a).  A possible optical counterpart
is a faint blue (in infrared colors) star, at $M_J\simeq1.7$ when the
X-ray source was faint. Heinke et al.\ (2003a) note that the X-ray
spectrum when the source is luminous 
is like those of the luminous sources in NGC\,6624 and NGC\,6712, and suggest that the
source is an ultra-compact binary (see Table 8.1). If that is the case,
its high optical flux is surprising. Wijnands et al.\ (2005) find that
the spectrum in quiescence (near $10^{33}$\ergsec) is dominated by
a hard power-law component.

{\it Terzan 6}. Extended studies of Terzan 6 with RXTE show that the
transient X-ray source in this cluster has fairly frequent outbursts,
on average every 140 days (in 't Zand et al.\ 2003).  An X-ray
position, derived from a Chandra observation, and an improved position
for the center of the cluster, found with ESO NTT observations, show
that the X-ray source is close to the cluster center.  The RXTE
observations provide an upper limit to the change in the orbital
period: $|\dot P/P|<3\times10^{-8}$~yr$^{-1}$.

{\it Liller\,1}. The Rapid Burster in Liller\,1 is a recurrent
transient. It shows a bewildering variety of X-ray behavior. When
discovered in 1976 (Lewin et al.\ 1976), it emitted X-rays largely in
the form of very frequent bursts (which were later called type II
bursts).  The average burst rate was in excess of $10^3$ per day; this
gave the source its name. There is an approximate linear relation
between the burst fluence and the waiting time to the next burst
(i.e.\ the mechanism is like that of a relaxation oscillator). These
rapid bursts are the result of spasmodic accretion.  Type II bursts
have been observed that lasted up to ten minutes with a corresponding
waiting time to the next burst of $\sim$1 h.  At times (early in an
outburst which typically lasts several weeks), for periods of many
days, the Rapid Burster behaved like a normal \lmxb\ (i.e., persistent
emission, but no type II bursts). The Rapid Burster also produces the
thermonuclear, type I, bursts (Hoffman, Marshall \& Lewin, 1978).  A
review of this remarkable source is given by Lewin, Van Paradijs \& Taam 
(1993); see also \S2.9.5\mk. \nocite{hml78}\nocite{lpt95} An accurate Chandra position of
the Rapid Burster (Homer et al.\ 2001b) coincides with the radio
counterpart (Moore et al.\ 2000). The Einstein position of the Rapid
Burster (Hertz \& Grindlay 1983) is not compatible with the radio
counterpart and with the Chandra position. However, it does coincide
with one of three low-luminosity sources also detected with Chandra.
Perhaps the low-luminosity source was more luminous at the time of the Einstein
observations. On the basis of their luminosities, the low-luminosity sources
are probably low-mass X-ray binaries in quiescence (Homer et al.\
2001b).

\section{The \gc\ sources outside the Galaxy}

In this section, we discuss the very luminous \gc\ X-ray sources observed
in galaxies other than our own. The observations we discuss were all
done with Chandra, except for the ROSAT observations of M\,31.  Some
of the sources were already detected with ROSAT, but the positional
accurracy of Chandra allows more secure identifications with globular
clusters. Table 8.3 gives an overview of the observations reported so
far.  The lowest detectable luminosities vary strongly between
galaxies.  With the exception of M31 and NGC\,5128, however, we are
always talking about very luminous sources (the tip of the
iceberg). In addition to the sources discussed in this chapter,
sources in many other globular clusters associated with other galaxies
have been observed but not (yet) recognized as such, e.g.\ because the
required optical cluster studies are not available (see \S12 and
Table 12.1).

\begin{table*}
\begin{tabular}{lr@{\hspace{0.2cm}}r@{\hspace{0.1cm}}rlr@{\hspace{0.2cm}}r@{\hspace{0.1cm}}rlr@{\hspace{0.2cm}}r} 
 &     &     &       &   &        \multicolumn{4}{l}{HST--FOV} \\
galaxy & $X$ & $X_g$ & N & $S_N$ & $X$ & $X_g$ & N & $S_N$ &
 $L_l$ & $L_u$\\ 
NGC\,720\cite{jsbg03} & 42 & 12 & & 2.2\cite{kis97}
                 &   &  &    & &   38.6 & 40.0 \\ 
NGC\,1316\cite{kf03} & 81 & 5 & &  1.7\cite{gamm01} &
                       &   & & 0.9\cite{grid01} & 37.3 & 39.3 \\ 
NGC\,1399\cite{alm01} & 214 &    & 6450 & 5.1\cite{drg+03} 
                 & 45  & 32 &  678 &             & 37.7 \\ 
NGC\,1407\cite{whi02} & 160 & 88 & & 4.0\cite{phb+97} \\ 
NGC\,1553\cite{bsi01} & 49  & 2  & & 1.4\cite{az98}
                 &     &  & 1553 & 0.5\cite{kw01} & 38.3 & 39.3 \\ 
NGC\,3115\cite{kmzp03} & 90 & & & & 36 & 9 \\ 
NGC\,4365\cite{kmzp03} & 149 & & & 5.0\cite{kis97} & 44 & 18$^a$ & 660 & 2.1\cite{kw01} \\ 
NGC\,4472\cite{kmz02} & 135 &    & 5900 & 3.6\cite{rz01}
                 &  72 & 29 & 825  &             & 37.0 & \\ 
NGC\,4486\cite{jcf+04} & 174 & & 13450 & 14\cite{mhh94} 
                 &  98 & 60 &    &  & 37.2 & 39.0 \\ 
NGC\,4649\cite{rsi03} & 165 &  & & 6.9\cite{kis97}
                 & 40  & 20 & 497 & 1.4\cite{kw01}  \\ 
NGC\,4697\cite{sib01} &  80 & $>$8 & 1100 & 2.5\cite{kis97} 
                 &     &       &      &          & 37.7 & 39.4  \\ 
\\
M\,31\cite{shp+97}    & 353 & 27 & 500 &  1.2\cite{bh01} & & & & & 35.5 & 38.3 \\
M\,31\cite{skg+02}          &    &     &  &          
                 &  90 & 28 & & & 35.3 & 38.3\\
NGC\,4594\cite{skv+03} & 122 & 32 & 1900 & 2.1\cite{rz04}\\ 
NGC\,5128\cite{kkf+01} & 111 & 33 &      &  2.6\cite{har91}
                  &     & 29 &      &  & 36.2 \\ 
\end{tabular}
$^a$corrects number given in paper (Kundu private communication).
 \caption[o]{\it X-ray sources associated with globular clusters in
galaxies other than our own. For each galaxy we list the total number
of X-ray sources detected $X$, the number associated with globular clusters
$X_g$, and the number of globular clusters $N$ with specific frequency $S_N$
(Eq.\,\ref{eqsn});
and the same numbers again in a limited field-of-view (FOV; Hubble
Space Telescope observations, usually with the WFPC-2 
but with the ACS for M87\mk) where applicable.
We also list the logarithm
of the X-ray luminosity detection limit, $L_{l}$, and the luminosity, $L_u$, of the
most luminous cluster source, in \ergsec.
Numbers between {\rm [ ]} are references. \label{tgal}}
 \end{table*}

\begin{figure}[ht!]
\begin{center}
\resizebox{1.0\textwidth}{!}{\rotatebox{270}{\includegraphics{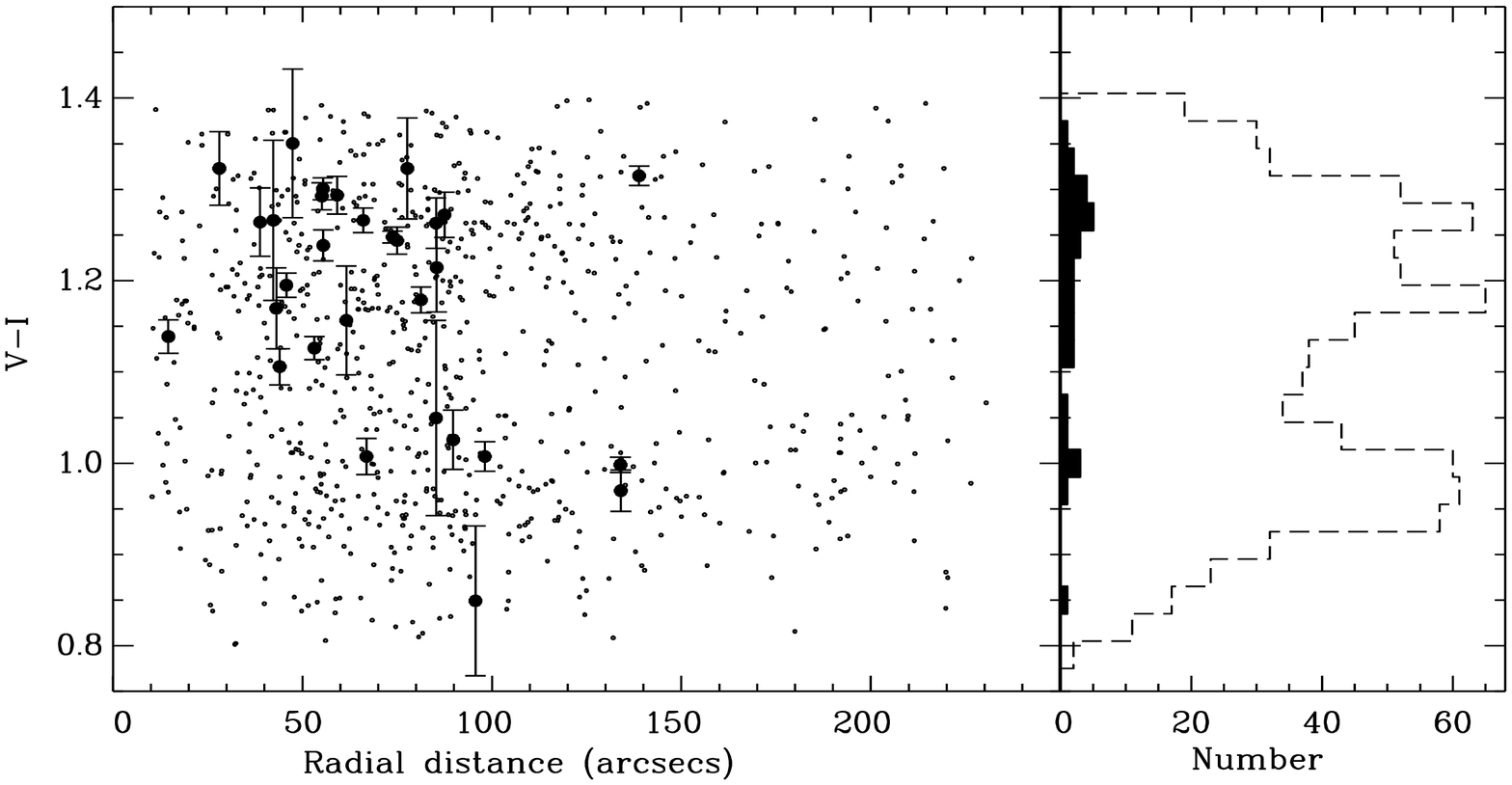}}}
{
\caption{\label{4472} Left: {\it V--I} colors of \gcs\ vs.\ distance
from the center of the elliptical \gal\ NGC\,4472.
\lmxb\--globular--cluster matches are indicated by filled circles.
Most of the \mk luminous \mbox{X-ray} sources are located in red \gcs. The optical color
distribution is shown on the right with a dashed line; notice the
bimodal distribution. The distribution of the \gcs\ that house the
luminous \mbox{X-ray} sources is also shown. Courtesy of Kundu, Maccarone \&
Zepf (2002).}  }
\end{center}
\end{figure}

The number of globular clusters varies widely between galaxies.
Precise numbers are difficult to determine: clusters are difficult
to detect against the \mk bright background of the central regions of a galaxy,
and the cluster distribution may extend beyond the observed area.
For example, globular clusters in NGC\,4697 have only been
identified in an annulus from 1.5 to 2.5 arcmin from the center (Fig.\ 8.5).
And even for nearby M\,31 ``the size of the globular cluster system is
embarrassingly uncertain'' (Barmby 2003). Estimates of the total
number are often based on an uncertain extrapolation of the measured bright
part of the globular cluster luminosity function and depend
on the availability of multi-color
images that go deep enough to probe a significant portion
of the luminosity function (Kundu, private communication).
In many galaxies the area in which positions of globular clusters
are known with sufficient accuracy for comparison with X-ray positions
is limited by the field-of-view of HST-WFPC2 observations:
an example is seen in Figure 8.4.

The number $N$ of globular clusters of a galaxy is sometimes scaled
to the total luminosity of the galaxy (derived from absolute
magnitude $M_V$), as a specific frequency
$S_N$, defined as (Harris \&\ van den Bergh 1981):
\begin{equation}\log S_N = \log N + 0.4(M_V+15)\label{eqsn}\end{equation}
A `local' specific frequency is often defined for the field-of-view
of the HST-WFPC2. The uncertainties in the total number of globular
clusters are reflected in large uncertainties of the specific frequencies,
and the uncertainty in the distance adds to this.
For example, values for NGC\,1553 range from 1.22$\pm$0.27 to
2.3$\pm$0.5 (Bridges \&\ Hanes 1990, Kissler-Patig 1997).
Specific frequencies (most are meant to be global) were \mk compiled by
Harris (1991),  Kissler-Patig (1997), and Ashman \&\ Zepf (1998).  
Local specific frequencies of \gcs\ have been measured in the inner 
region of 60 \gals\ (Kundu \& Whitmore, 2001a,b).

Many elliptical galaxies, and especially those in the center of
clusters of galaxies,  have large numbers of globular clusters
(Harris 1991; Ashman \& Zepf 1998). Per unit mass,
most ellipticals have about twice as many
\gcs\ as \mk spirals (Zepf \& Ashman 1993, 1998). The \gc\ populations 
in most elliptical \gals\ show a bimodal distribution in optical 
colors (Figure 8.3). Most of this is due to
differences in metallicity, but differences in age may also play
a role. Metal-poor
clusters are bluer than metal-rich clusters of the same age;
at the same metallicity, old clusters are redder than young ones.
It has been suggested that the blue metal-poor \gcs\ were formed at the 
proto-galactic epoch, and that the red metal-rich \gcs\ resulted 
from \mk later starbursts, e.g.\ as a consequence of the mergers that produce the
\gals\ that we observe today (Ashman \& Zepf, 1992; Zepf \& Ashman,
1993; for other possibilities see the review by West et al.\ 2004). 
However, to date there is no convincing evidence for difference in ages 
of red and blue subsystems (e.g., Puzia et al.\ 2002, Cohen et al.\
2003, C\^ot\'e 1999,  C\^ot\'e et al.\ 2002).

\subsection{Elliptical galaxies}

The most luminous X-ray sources in a galaxy are high-mass X-ray
binaries, supernova-remnants, and low-mass X-ray binaries.  Since
elliptical galaxies do not house young stellar populations, virtually
all luminous X-ray sources in them will be low-mass X-ray
binaries. Table\,\ref{tgal} provides an overview of the references and
results; some additional remarks for individual galaxies follow. We
will discuss the X-ray luminosity functions of the globular cluster
systems and the reported breaks in some of them in \S8.3.4.

\begin{figure}[ht!]
\begin{center}
\resizebox{0.7\textwidth}{!}{\rotatebox{270}{\includegraphics{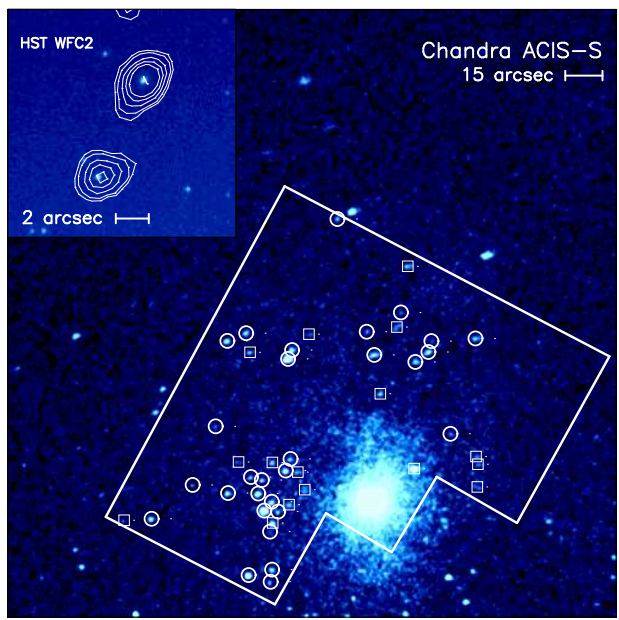}}}
{
\caption{\label{1399} The 0.3--10 keV Chandra image of NGC\,1399 centered 
on an HST pointing, smoothed with a Gaussian of about 0.8$''$. The
white line marks the HST/WFPC2 FOV. The circles show the X-ray source
positions that are associated with \gcs. The squares are the remaining
sources. All 45 sources are marked; 38 have a significance in excess
of 3$\sigma$. The top left image is an example of the Chandra
contours overlaid on the HST field. Courtesy Angelini et al.\ (2001).
} }
\end{center}
\end{figure}

{\it NGC\,1399} is a giant elliptical galaxy in the center of the 
Fornax Cluster at 20.5\,Mpc. 
A large fraction of the 2-10 keV \mbox{X-ray} emission
in an $8'$x$8'$ region is resolved into 214 discrete sources,
including many background sources.
32 are in globular clusters (see Figure\,\ref{1399}). Many of
the \gc\ sources have super-Eddington luminosities (for an accreting
neutron star), and their average luminosity is higher than that of the
sources not associated with \gcs. The most luminous source in a \gc\ \mk 
has an ultra-soft spectrum such as \mk seen in the high state of
black-hole binaries. This may indicate that some of the
most luminous sources are binaries with an accreting
black-hole, rather than
conglomerates of less luminous neutron-star binaries (Angelini et 
al.\ 2001).

Dirsch et al.\ (2003) find that ``within 7$'$ the \SN\ of the 
blue clusters alone is a factor $\sim$3 larger than for the red ones. 
Outside this radius, both populations have the same high local \SN'',
listed in Table\,8.3.

{\it NGC\,4697}. In this galaxy, most of the \mbox{X-ray} emission is
from point sources. The central source, with
$L_X=8\times10^{38}$~\ergsec, may be an active nucleus and/or multiple
\lmxbs\ (Sarazin et al.\ 2000, 2001). \\

\begin{figure}[ht!]
\begin{center}
\resizebox{0.7\textwidth}{!}{\includegraphics{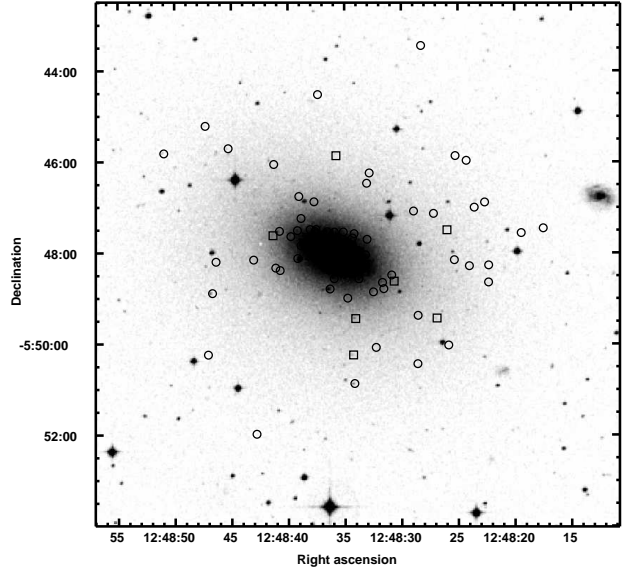}}
{
\caption{\label{4697} Digital Sky Survey \mk optical image of NGC\,4697. The circles show
the positions of the \mbox{\mbox{X-ray}} sources detected with \chandra. The
squares indicate \mbox{X-ray} sources in known \gcs.  One should note that
\gcs\ have only been identified in this galaxy in an annulus from 1.5
- 2.5 arcmin from the center. This figure was kindly provided by Craig
Sarazin. It is adapted from Figure 3 of Sarazin, Irwin \& Bregman
(2001).
}
}
\end{center}
\end{figure}

{\it NGC\,4472} is a giant elliptical galaxy. In the inner regions of
the galaxy it has been shown that metal-rich red \gcs\ are about 3
times more likely to host a very luminous \lmxb\ than the blue
metal-poor ones (Figure 8.3).  The \mbox{X-ray} \lum\ does not depend
significantly on the properties of the host \gc\ (Kundu et al.\ 2002).

{\it NGC\,4365}, in the Virgo cluster, is one of a few early-type \gals\ 
whose globular clusters do not have a bi-modal color distribution in $V$$-$$I$
(but it does in infrared colours, Puzia et al.\ 2002).
Kundu et al.\ (2003) find that the presence of very luminous \lmxbs\ is
correlated with metallicity, but not with cluster age. 
The \lmxb\ fraction
per unit mass of the \gcs\ is $\sim$$10^{-7}$ $\Msun^{-1}$.
In contrast, 
Sivakoff et al.\ (2003) 
find that within the sample of IR-bright
\gcs\ studied by Puzia et al.\ (2002), the metal-rich, intermediate-age \gcs,
are four times as likely to contain \lmxbs\ than the
old \gcs\ (with an uncertainty of a factor of two). The \lum\ function
is a power-law with a cutoff at $\sim$(0.9-2.0)$\times$10$^{39}$ \ergsec, 
much higher than the cutoff measured for other ellipticals.

{\it NGC\,3115} has a distinct bimodal color distribution of the \gcs.  
The metal-poor blue and the metal-rich red \gcs\ are both
$\sim$12 Gyr old (Puzia et al.\ 2002). There are roughly equal numbers
of red and blue \gcs\ in the WFPC2 image. Kundu et al.\ (2003)
find that the red \gcs\ are the preferred sites for \lmxb\ formation,
largely as a consequence of their higher metallicity.

{\it NGC\,1407}. White (2002) reported that about 90\% of the 160
detected \lmxbs\ have \mbox{X-ray} \lums\ which exceed the Eddington
limit for neutron stars.  He suggests that many may be \bh\ binaries
(rather than multiple neutron-star binaries within individual \gcs),
since 45\% do not reside in \gcs. To date (September 2004), these results
have not yet been published in a refereed journal.

{\it NGC\,1553} is an S0 galaxy. 30\%\ of the
emission in the 0.3--10 keV band and 60\% of the 
emission in the 2.0--10 keV band is resolved into discrete sources (Blanton, Sarazin \& Irwin, 2001).
 
Kissler-Patig (1997) lists a global specific frequency of 2.3 $\pm$ 0.5, 
higher than the value listed in Table\,8.3.

{\it NGC\,4649} (M60) is a bright elliptical \gal. It was observed by
Randall, Sarazin \& Irwin (2003); for details see Table 8.3.

{\it NGC\,1316 (Fornax A)} is a disturbed
elliptical radio \gal\ with many tidal tails. Several mergers must
have occurred over the past 2~Gyr (see Kim \& Fabbiano 2003, and
references therein). One of the 5 globular cluster sources is
super-soft. For an adopted
distance of 18.6 Mpc,  35\%\ of the sources are above the Eddington limit of
a 1.4\Msun\ neutron star (Kim \&
Fabbiano, 2003). The luminosity function is well represented by a
power law with a slope of $-$1.3. 

{\it NGC\,720}. 3 of the 12 globular cluster sources have 
X-ray luminosities in excess of 10$^{39}$ \ergsec\ (at 35 Mpc). It is possible that this galaxy is much closer, and that none of the
sources are ultra-luminous (Jeltema et al.\ 2003).

{\it NGC\,4486} (M\,87) is a giant elliptical galaxy near the center of
the Virgo cluster, and has the most populous globular cluster system in the local
supercluster. More luminous, redder and denser
clusters are more likely to harbor a luminous X-ray source. 
Metal-rich red globular clusters are about three times more likely to host
a luminous LMXB than the blue metal-poor ones; the trend with central
density gives strong evidence that encounter rates are important in
forming LMXBs in globular clusters. The trend with luminosity can arise as a consequence
of the fact that more luminous clusters have higher encounter rates.  The
X-ray luminosity functions of both globular cluster-LMXBs and non-globular
cluster LMXBs are well described by single power laws with an upper cutoff
at $\sim 10^{39}$ erg s$^{-1}$ (Jord\'an et al.\ 2004).

\subsection{Spiral Galaxies}

Plenty of spirals have been observed with \chandra, but there is very
little information on the optical identifications.  Globular clusters
are hard to find because of the patchy extinction.  There is also the
difficulty of subtracting the diffuse light of the \gal.  These
problems are exacerbated for nearly face-on spirals.  In addition to the
low-mass X-ray binaries, a spiral galaxy also hosts high-mass X-ray
binaries and supernova remnants among the luminous X-ray sources.

{\it M31 (Andromeda Nebula)}. The apparent size of M\,31 is so big that
only ROSAT has studied the whole (Magnier et al.\ 1992; Supper et al.\ 1997).
Di Stefano et al.\ (2002) have conducted \chandra\ observations of 
$\sim$2560 arcmin$^2$ in four different areas so as to be representative
of the whole. About one third of the 90 \chandra\
sources have \lums\ (0.5--7 keV) in excess of $10^{37}$ \ergsec;
the most luminous source is probably associated with the
\gc\ Bo~375. Its \lum\ (0.5--2.4 keV) varied between
$\sim$2$\times$10$^{38}$ and $\sim$5$\times$10$^{38}$ \ergsec.
Supper et al.\ (1997) reported regular variations of $\sim$50\% on a
timescale of $\sim$16 hours.  A similar percentage variability was
found in the 500 day X-ray light curves of two other highly luminous \gcs\
in M31, Bo~82 and Bo~86 (Di Stefano et al.\ 2002). 
Some of the more luminous \gc\ X-ray sources could be multiple sources.

\begin{figure}
\centerline{\psfig{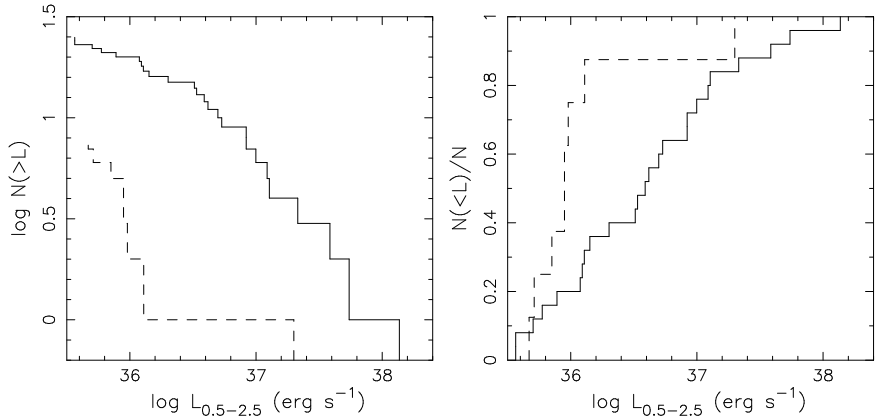}}

\caption[o]{\it Left: Comparison of the cumulative (from
high luminosities downward) X-ray luminosity
distributions of globular clusters in the Milky Way (dashed curve) and 
M\,31 (solid curve). After Di Stefano et al.\ (2002). Right: Normalized
cumulative (from low-luminosities upward)
X-ray luminosity function for sources with $L_x>10^{35.5}$\ergsec\ 
in the Milky Way (dashed curve) and M\,31 (solid curve). The Chandra
luminosities given by Di Stefano et al.\ (2002) were multiplied by 0.46
to convert them to the
energy range of the ROSAT data from Verbunt et al.\ (1995). The probability 
that the normalized distributions are the same is 0.03.
\label{verbfks}}
\end{figure}

It has been stated on the basis of different data sets that the
X-ray luminosity function of globular cluster X-ray sources
is different in M\,31 than in the Milky Way (Van Speybroeck et al.\
1979 on the basis of Einstein data; Di Stefano et al.\ 2002), and
that it is the same (Supper et al.\ 1997). In Figure~\ref{verbfks} we show the
normalized cumulative distributions for clusters in the Milky Way 
and in M\,31. The distributions look different, but a Kolmogorov-Smirnov
test shows that there is a non-negligible probability, 0.03, that the
difference is due to chance. It is therefore possible that
the extent to higher luminosities in M\,31 is due to the larger number
of X-ray sources (and of globular clusters).

{\it M104 (NGC\,4594, Sombrero galaxy)} is an Sa galaxy at a
distance of $\sim$8.9 Mpc. Only optically
bright \gcs\ house the luminous \lmxbs\ detected with \chandra\ (Di
Stefano et al.\ 2003). 
The majority of the sources with \lums\ in excess of $10^{38}$
\ergsec\ are located in \gcs. The \lum\ function of X-ray sources in
the \gcs\ has a cut-off near the Eddington limit for a 1.4\Msun\ neutron star. 
One \gc\ houses a super-soft source (see \S11).
There is a connection
between metal-rich, red \gcs\ and the X-ray sources. However,
the most luminous X-ray sources are equally likely to be located in
metal-poor \gcs\ with lower optical luminosities. The optically
brightest blue \gcs\ do not seem to house very luminous X-ray sources.

\subsection{NGC\,5128 - Cen\,A}
This galaxy is probably the result of mergers, and
consequently it is somewhat like a mixture between
an elliptical and a spiral. Four X-ray sources
outside the WFPC2 FOV are coincident with \gcs\ (Kraft et al.\
2001; Minniti et al.\ 2004).  70\%\ of the globular cluster sources have
\lums\ in excess of 10$^{37}$ \ergsec. There is no indication that any
of them are \bh\ binaries. The \gc\ X-ray sources are preferentially
found in massive \gcs. Most of the \gcs\ which harbor a luminous X-ray
source have red colors (metal-rich). NGC\,5128 is at a low galactic
latitude; there is a lot of foreground extinction. This makes it
difficult to get reliable optical data on \gcs.

\subsection{Comparison and interpretation}
Many galaxies contain a substantially larger number of luminous X-ray
sources in globular clusters than our own galaxy (compare
Tables\,\ref{tbright} and \ref{tgal}).  This can be explained by their
larger numbers of globular clusters.  The fraction of globular
clusters that contains a luminous X-ray source is roughly constant
between galaxies, as is the number of X-ray sources in clusters scaled
on cluster luminosity or mass ($2\times10^{-7} {L_{\odot,I}}^{-1}$ for
$L_x>3\times10^{37}$\ergsec, Sarazin et al.\ 2003, Kundu et al.\
2003).  Similarly, the larger number of globular cluster X-ray sources
in M\,31 compared to the Milky Way may be explained by the larger
number of clusters (Supper et al.\ 1997, Di Stefano et al.\ 2003).
Several authors reported a knee near the Eddington \lum\ for an
accreting neutron star in the luminosity functions of ellipticals
(Sarazin et al.\ 2000, 2001 for NGC\,4697; Kundu et al.\ 2002 for
NGC\,4472; Blanton et al.\ 2001 for NGC\,1553, and Randall et al.\
2003 for NGC\,4649).  However, Kim \&\ Fabbiano (2004), who corrected
the Chandra data for incompleteness, have shown that the luminosity
functions for each of the observed elliptical galaxies can be fit with
one power law; two power laws do not improve the fit in a significant
way.  It is interesting to note that even though no breaks in the
individual luminosity functions are significant, if the luminosity
functions of all observed ellipticals (containing a total of about 985
point-like sources) are added, a broken power-law fit is a better fit
than a single power law; the break is near $5\times10^{38}$\ergsec\
(Kim \&\ Fabbiano, 2004; \S12.4.3).

Clearly, a large number of LMXBs have luminosities substantially above
the Eddington luminosity of an accreting neutron star. In analogy with
the luminosity distribution in the Milky Way (Grimm et al.\ 2002),
this suggests that many of these sources may be accreting black holes.
This suggestion is supported in some cases by the X-ray spectrum,
which shows the soft signature of an accreting black hole (e.g.\
Angelini et al.\ 2001). The fact that a very luminous accreting black
hole is not found in the globular clusters of the Milky Way is
probably due to the small number of cluster sources.

An alternative explanation for the LMXBs with luminosities
substantially above the Eddington luminosity of an accreting neutron
star is provided by Bildsten \&\ Deloye (2004), who note that the
Eddington limit for hydrogen-poor gas is higher. They show that the
high luminosities can be explained by invoking binaries in which a
helium or carbon/oxygen white dwarf of 0.04-0.08 \Msun\ transfers mass
to a neutron star, at orbital periods of 5-10 minutes.  From the
evolution of such binaries, driven by gravitational radiation, to
longer periods and lower mass-transfer rates the expected luminosity
function can be computed, and is found to be compatible with the
observed luminosity function.

The X-ray sources are found preferably in optically bright clusters
(Angelini et al.\ 2001).
This could be explained as a scaling with mass (Kundu et al.\ 2002,
Sarazin et al.\ 2003).
We suggest, however, that the scaling with mass is a proxy for the
scaling with the collision number, caused by the strong correlation
between mass and collision number. 
In the Milky Way, the probability of a cluster to contain a luminous
X-ray source scales better with the collision number than with
the mass (Verbunt \&\ Hut 1987; Pooley et al.\ 2003). 

\begin{figure}
\centerline{\psfig{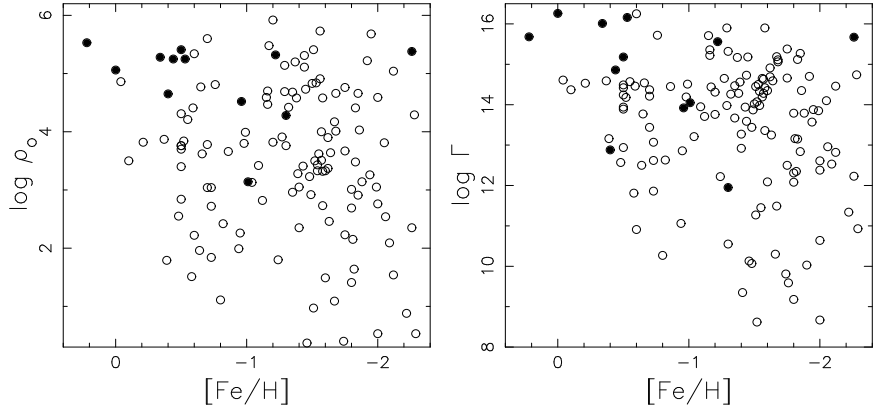}}
\caption[o]{\it Left: central density of globular clusters
in the Milky Way as a function of metallicity. Filled circles indicate
globular clusters with a luminous X-ray source. Even at the same density
there is a preference for high-metallicity clusters. After
Bellazzini et al.\ (1995). Right: the preference for high-metallicity
clusters persists in a plot of collision number (Eqs.\ 8.5 and 8.6) as a function of metallicity.
\label{verbffeg}}
\end{figure}

In many galaxies, luminous X-ray sources are found preferably in red,
metal-rich clusters.  Bellazzini et al.\ (1995) demonstrated this for
the Milky Way (see Figure\,\ref{verbffeg}) and less conclusively for
M\,31. Di Stefano et al.\ (2003) find in their sample of M\,31
clusters that the probability that a cluster contains an X-ray source
is not strongly correlated with metallicity. Kundu et al.\ (2002)
find that a red cluster in NGC\,4472 has a 3 times higher probability
of hosting a luminous X-ray source than a blue cluster. A similar result
is found for NGC\,4365 by Sarazin et al.\ (2003), and for NGC\,3115
by Kundu et al.\ (2003). We consider four suggested explanations. 
First, if metal-rich clusters are younger, they contain main-sequence 
stars of higher mass, which are thought to be more efficient in forming 
an X-ray binary (Davies \&\ Hansen 1998). In NGC\,4365 such a young 
population is indeed \mk present, but it does not show an increased 
formation rate of X-ray sources (Kundu et al.\ 2003). Also, the preference
for metal-rich clusters is observed in the Milky Way and in NGC\,3115, 
where all globular clusters are old. These results show that 
metallicity, not age, must explain the preference of X-ray sources for 
red clusters (Kundu et al.\ 2003).
Second, a higher X-ray luminosity at higher metallicity would produce
a preference for metal-rich clusters in a flux-limited sample.
Various models have been suggested to produce higher X-ray luminosities
in binaries with a donor of higher metallicity (e.g.\ Bellazzini et al.\ 
1995, Maccarone et al.\ 2004). However, X-ray sources in
metal-rich clusters are not observed to be more luminous than those in 
metal-poor clusters in M\,31 (Verbunt et al.\ 1984) or, with
less statistical constraint, in NGC\,4472 (Maccarone et al.\ 2003). 
Third, Grindlay (1987) suggests that metal-rich clusters have
a flatter initial mass function (and hence more neutron stars).
However, such a dependence is not observed in the Milky Way
(Piotto \&\ Zoccali 1999).
Finally, Bellazzini et al.\ (1995) suggest that the longer life times
and larger radii of metal-rich stars enhance their capture rate; the
capture probability is proportional to radius (see Eq.\,\ref{gamma}
below), and it must be doubted that the small difference in radii has
sufficient effect to explain the observations (Maccarone et al.\ 2004). 
It is fair to say that the connection between metallicity and the
occurrence of \lmxbs\ in \gcs\ is not yet well understood.

There is a tendency for X-ray sources in metal-rich globular
clusters to have softer X-ray spectra (M\,31: Irwin \&\ Bregman 1999,
NGC\,4472: Maccarone et al.\ 2003).

\subsection{Comparison between field and cluster sources\label{secfield}}

The X-ray luminosity function of sources in globular clusters
is not very different from that of the sources outside globular
clusters (Maccarone et al.\ 2003, Sarazin et al.\ 2003).
The spatial distribution of X-ray sources outside globular clusters
in elliptical galaxies is similar to that of the globular cluster
sources. 
In elliptical galaxies, globular clusters often harbor a very large
fraction of all X-ray sources (Table\,\ref{tgal}). This has raised
the suggestion that {\em all} X-ray sources in elliptical galaxies
originate in globular clusters (White et al.\ 2002).
The field sources then could have been ejected from a cluster,
or originate in a cluster that was later destroyed by the galactic tidal 
field. The demand that a cluster lives long enough to form X-ray binaries,
and short enough not to be around now, requires fine tuning.
Thus, the ejection hypothesis may be more probable.

This would suggest that a large number of globular clusters translates
into a large number of X-ray sources, both in the clusters and (due to
ejection) outside them. The fraction of X-ray sources in globular
clusters would then be similar for different galaxies.  In the Milky
Way and in M\,31 there are about 10 luminous low-mass X-ray binaries
in the disk for each one in a globular cluster.  In elliptical
galaxies, there is of order 1 low-mass X-ray binary outside clusters
for each one in them (see Table\,\ref{tgal}). This indicates that the
majority of the disk sources in the Milky Way and M\,31, and by
extension in spiral galaxies in general, are formed in the disk;
although as noted in the Introduction some individual systems may have
escaped from globular clusters. [One should note that the HST field of
view is much smaller than that of Chandra. Therefore, in comparing the
number of X-ray sources associated with globular clusters with those
not located in globular clusters (Table 8.3), in all those cases where
HST data were needed to identify the clusters, one can only consider
the X-ray sources which are detected in the regions observed with
HST.]

For elliptical galaxies the case is less clear. Using optical
luminosities of the galaxies and the specific globular cluster
frequencies, White, Sarazin \& Kulkarni (2002) reported that they
found evidence that the sum of the X-ray luminosities of all X-ray
sources in ellipticals scales approximately with the number of
globular clusters, and they conclude that this indicates that the
population outside clusters is formed in the clusters. However, the
uncertainties in the specific frequencies may be substantially larger
than the values used by these authors, and that makes it difficult to
quantify their findings. Kim and Fabbiano (2004) have made a similar
study, and caution about the above interpretation.

The fraction of low-mass X-ray binaries in clusters ranges from about
20 to 70 \% in ellipticals (see Table 8.3). This suggests, in our
opinion, that globular clusters alone are not responsible for all
low-mass X-ray binaries.  In systems with
small numbers, the total luminosity can be affected by just a couple
of very luminous sources; the number of
sources may therefore be a better estimator for the population size than the
integrated X-ray luminosity. Clearly, the origin of low-mass X-ray
binaries in elliptical galaxies deserves more study.

If the majority of those luminous \lmxbs\ in
\elgals\ not located in \gcs\ are primordial, their luminosities
could not have been constant throughout their lifetimes (because the
product of age and the required mass-transfer rate would exceed the donor
mass). There are two
ways out of this lifetime problem: (i) they are not primordial but
they were formed in \gcs,\ and somehow released into the field, or
(ii) the majority of them are transients with a low duty cycle (see
Piro \& Bildsten 2002). If the latter is the case, follow-up \mk
observations with \chandra\ will be able to reveal the variability if a
sufficient number of them have outbursts that last only a few years
and not much longer.
We may add a third solution, which is that (iii) systems formed from
primordial binaries will emerge from their early evolution as
neutron stars or black holes with detached main-sequence companions.
How long it takes for the binary to turn into an X-ray source
then depends on the time required for the orbit to
shrink due to loss of angular momentum, or for the donor to
expand into a giant after completing its main-sequence evolution
(for reviews see Verbunt 1993 and Chapter 16 by 
Tauris and Van den Heuvel).
It may be noted that binaries formed in a globular cluster
may also go through a long-lived detached phase (Grindlay 1988).

\section{Low-luminosity X-ray sources\label{vlsecfs}}

\begin{figure}
\centerline{
\parbox[b]{6.5cm}{\psfig{figure=verbftuc.ps,width=6.5cm,clip=t}}
\parbox[b]{6.5cm}{\psfig{figure=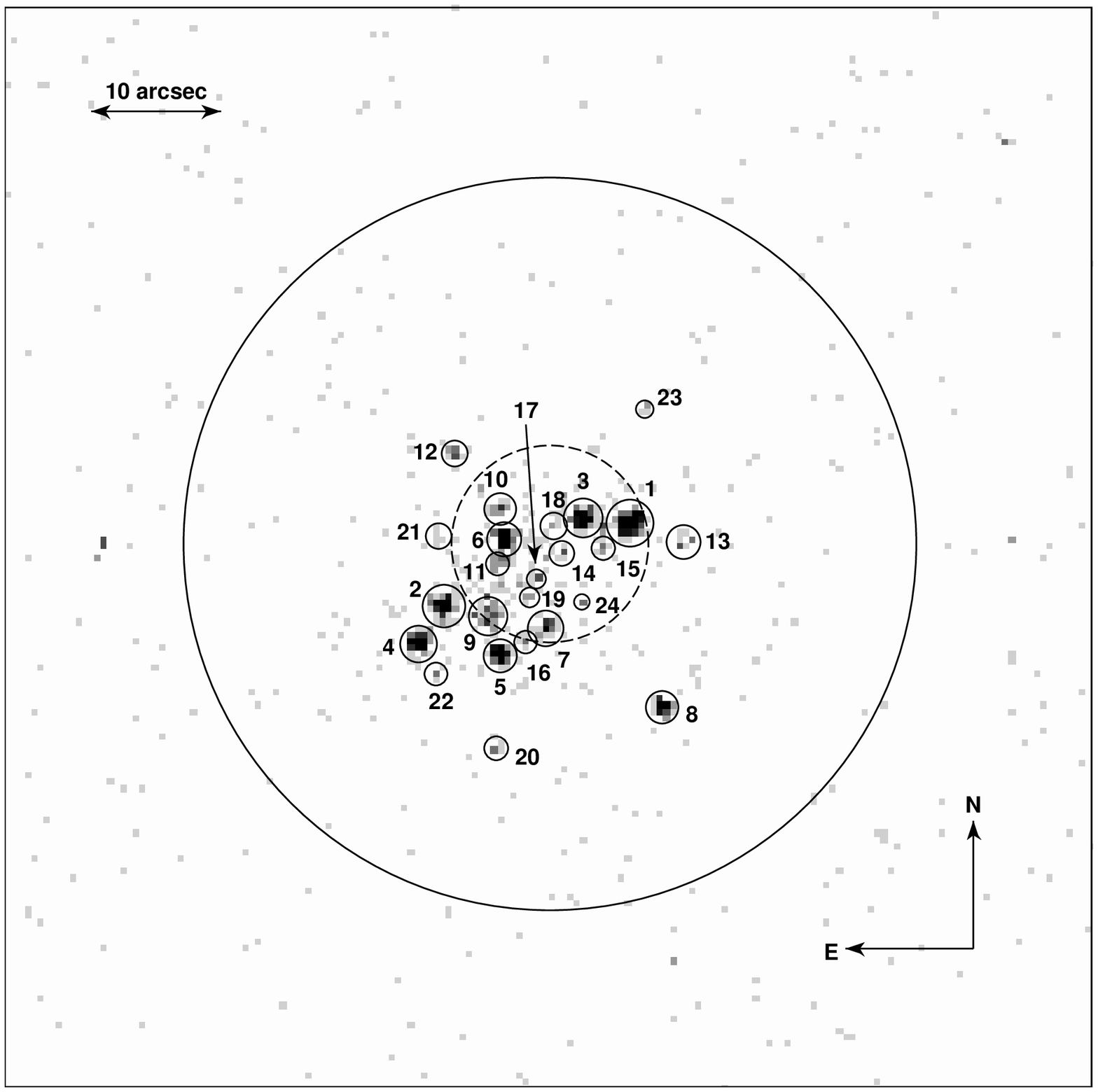,width=6.5cm,clip=t}}
}
\caption[o]{\it Left: The impact of Chandra on the study of low-luminosity
X-ray sources in globular clusters is well illustrated by the observations of
47 Tuc. The grey scale of the smoothed ROSAT-HRI countrate indicates the
resolution obtained with this instrument (Verbunt \&\ Hasinger 1998).
The circle indicates the position (1$\sigma$ region) of the
single Einstein source (Hertz \&\ Grindlay 1983), 
squares indicate the positions of the five ROSAT
sources in this region, filled circles are 39 Chandra positions
(Grindlay et al.\ 2001a). 
Right: Projected distribution of X-ray 
sources in the globular cluster NGC\,6440. The dashed and solid
lines indicate the core and half-mass radii, respectively.
From Pooley et al.\ (2002b). In the case of 47\,Tuc, each ROSAT source 
corresponds to one Chandra source; in the case of NGC\,6440, two
sources previously found by ROSAT
are both resolved into multiple sources. \label{verbfa}}
\end{figure}

As already mentioned, a limited number of low-luminosity sources has been detected
with Chandra in several clusters which contain a luminous X-ray source.  The
presence of such a source limits the sensitivity with which
low-luminosity sources can be detected, because of the wings of the
point spread function. The sharp ($<1''$) images and high sensitivity
of the Chandra observations is best used in clusters which do not
contain a luminous source. Such observations show that the central
regions of several globular clusters contain dozens of sources.
As a typical example, the distribution of the
sources in NGC\,6440 is concentrated towards the cluster center; while
it spreads beyond the core radius, it is fully contained within the
half-mass radius (Figure\,\ref{verbfa}).  
From this spatial distribution alone, it can be
safely asserted that almost all sources detected are related to the
globular cluster.  In clusters with large apparent
core radii and/or half-mass radii, a large fraction 
of the detected sources may be fore- or background
sources; an example is $\omega$\,Centauri.

In trying to determine the nature of all these X-ray sources,
we may be guided by our knowledge from previous satellites, in particular
ROSAT. Such guidance allows us to make a preliminary classification of a
source based on its X-ray flux and spectrum.
If a secure optical counterpart is found -- which thanks to the accurate 
source positions of Chandra is often the case whenever
sufficiently deep HST observations are available -- the classification
of a source can be further based on its optical spectrum, and on the
ratio of the X-ray and optical fluxes.
A secure classification can also be found if the position of
a radio pulsar coincides with that of an X-ray source: radio
and X-ray positions are so accurate that the probability of a
chance coincidence is virtually negligible for these rare objects.

Our discussion of the low-luminosity sources proceeds through the various classes
illustrated in Figure\,\ref{verbfb}, viz.\ low-luminosity low-mass X-ray binaries,
recycled radio pulsars, cataclysmic variables, and magnetically active
close binaries.
An overview of published Chandra observations of low-luminosity sources
in globular clusters is given in Table\,\ref{verbta}.

\subsection{Low-luminosity low-mass X-ray binaries \label{secflmx}}

We consider a low-luminosity low-mass X-ray binary with a neutron
star, \lmns, securely classified when its luminosity
is high enough ($\Lx \gtap 10^{32}$~\ergsec) and its X-ray
spectrum is soft (black body color temperature about 0.1 to 0.3 keV).
The reason for this is that most soft X-ray transients in the
galactic disk have these properties when they contain a neutron star.  
Their quiescent X-ray spectra have been roughly described as Planck
spectra with a temperature of about 0.3\,keV (Verbunt et al.\ 1994),
but more correctly should be fitted with model spectra of neutron star
atmospheres as have been computed by e.g.\ Rajagopal \&\ Romani (1996)
and Zavlin et al.\ (1996).  For quiescent transients in the disk, such
fits give effective temperatures of 0.1--0.2\,keV and neutron star
radii of roughly 10\,km (Rutledge et al.\ 1999).
\nocite{vpe84}\nocite{pvsa87}\nocite{vbj+94}\nocite{rr96}\nocite{zps96}
\nocite{rbb+99}
The situation is more problematic if a transient in
quiescence has a power-law spectrum and a \lum\ in the range $10^{31}
- 10^{34}$ \ergsec.  In that case, the system could be either a \lmns\
or a low-mass X-ray binary with a black hole, 
\lmbh\ (see Tomsick et al.\ 2003, Wijnands et al.\ 2005).
A hard spectrum can also indicate a cataclysmic variable, as
may be the case for one or two sources in NGC\,6652 and Terzan\,1.
\nocite{tcf+03}\nocite{whp+05}

\begin{figure}
\centerline{
\parbox[b]{7.5cm}{\psfig{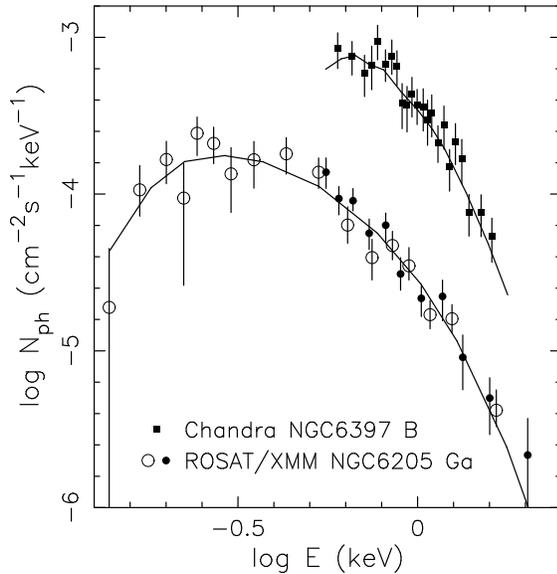}}
\parbox[b]{4.5cm}{\caption[o]{\it X-ray spectra of low-luminosity
X-ray binaries with neutron stars, as observed (i.e., not corrected
for interstellar absorption) with ROSAT and XMM for the source Ga in
NGC\,6205 (M\,28) (Gendre et al.\ 2003b; Verbunt 2001); and with
Chandra for the source B in NGC\,6397 (shifted upwards by 1 decade\mk; in 't
Zand, private communication; Grindlay et al.\ 2001b). The solid lines
show fits with models for hydrogen atmospheres of neutron stars.
\label{verbfspec}}
}
}
\end{figure}

Chandra and XMM are sensitive enough to detect
luminosities of $L_x\gtap10^{32}{\rm erg s}^{-1}$ in any 
cluster
that they observe, with sufficient counts to determine whether the
spectra are power laws or thermal (i.e., soft).  Sources for
which fits with neutron star atmosphere models have been shown to give
a good description of the X-ray spectrum include X7 in $\omega$\,Cen
(Rutledge et al.\ 2002, see also Gendre et al.\ 2003a), X5 and X7 in
47\,Tuc (Heinke et al.\ 2003b), B in NGC\,6397 (Grindlay et al.\
2001b), CX1 in NGC\,6440 (in 't Zand et al.\ 2001), and Ga in
NGC\,6205 (M\,13, Gendre et al.\ 2003b).
\nocite{rbb+02}\nocite{hgle03}\nocite{gbw03}\nocite{gbw03b}\nocite{ghe+01}
\nocite{zkp+01} Most of these sources were
already detected with ROSAT, being (among) the most luminous sources in
each cluster (the exception is CX1 in NGC\,6440).  As noted above, CX1
in NGC\,6440 is the transient, detected in the bright state in 1998
and 2001; whether the transient of 1971 was the same source cannot be
ascertained. This source supports our premiss that the more luminous
($L_x\gtap10^{32}$ \ergsec) among the low-luminosity soft sources are quiescent
accreting neutron stars.

Probable classifications as low-luminosity \lmns, based on
the ratio of soft to hard counts as detected with Chandra have been
suggested for 4 of the most luminous faint sources in NGC\,6440 (Pooley et al.\
2002b), and in Terzan\,5 (Heinke et al.\ 2003a).  Further probable
identifications are based on the luminosity of the sources: 3 low-luminosity
\lmns\ (in addition to the Rapid Burster) in Liller\,1 (Homer et al.\
2001b), 1 or 2 in NGC\,6652 (Heinke et al.\ 2001).
\nocite{plv+02}\nocite{heg+03}\nocite{heg01}\nocite{hdam01} We want
to point out, however, that it cannot be excluded that some of these are 
black-hole binaries. 

A low-mass X-ray binary with a black hole can have a 
much lower luminosity than a \lmns; as an example, for the transient A0620$-$00
in quiescence $\Lx \simeq 10^{30}$~\ergsec, much of which could even be
due to the donor in the binary (Verbunt 1996, Bildsten \&\ Rutledge
2000).
At such low luminosities, even Chandra or XMM observations cannot
provide a secure classification, and consequently we have no
information on the number of low-luminosity low-mass X-ray binaries
with a black-hole accretor.
\nocite{ver96}\nocite{br00}

So far, only two low-luminosity \lmns s in globular clusters have been identified optically,
one in 47\,Tuc and one in $\omega$\,Cen (Edmonds et al.\ 2002b, Haggard et
al.\ 2004).
\nocite{ehgg02}\nocite{hca+03}

\subsection{Millisecond pulsars \label{secmsp}}
 
Most identifications of X-ray sources in globular clusters
with recycled radio pulsars are based on positional coincidence.
The exceptions are the identifications of the pulsar in NGC\,6626 (M\,28),
which is based on the pulse period, and of pulsars in NGC\,6397 (XB) 
and in 47\,Tuc (W29/PSR\,W) which are based on their orbital periods.

The pulsar in M\,28 is the only one in a globular cluster which was identified
with an X-ray source before the Chandra observations. By 
comparing the on-pulse X-rays with the off-pulse X-rays, the
X-ray spectrum of the pulse could be isolated (Saito et al.\ 1997).
Chandra resolves the pulsar  from other cluster sources
and obtains a phase-averaged power law spectrum with photon index 1.2
(Becker et al.\ 2003).

\begin{figure}
\centerline{
\parbox[b]{7.5cm}{\psfig{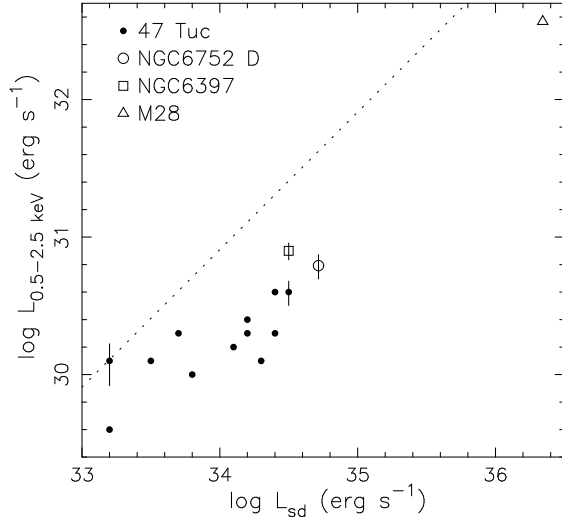}}
\parbox[b]{5.0cm}{\caption[o]{\it X-ray luminosities as a function of
spindown luminosities, $L_{sd}$, of radio pulsars in globular
clusters.  The unresolved pulsar pairs G/I and F/S in 47\,Tuc have
been omitted. The dotted line indicates
$L_{0.1-2.4\,keV}$$=$$10^{-3}L_{sd}$ (Verbunt et al.\ 1996), with a
small correction for the different X-ray energy range.  The cluster
pulsars lie below, but roughly parallel to this relation.  Notice M28
in the upper right hand corner.  Several 1$\sigma$ errors are shown;
these are computed from Poisson statistics of the detected number of
X-ray counts, and do not take into account uncertainties in the
spectral energy distribution and/or cluster distance. Data from
Grindlay et al.\ (2002), D'Amico et al.\ (2002), Becker et al.\ (2003)
and Possenti et al.\ (quoted in Bassa \&\ Stappers 2004).
\label{verbflxe}}
}
}
\end{figure}
\nocite{vkb+96}\nocite{gch+02}\nocite{dpf+02}\nocite{bsp+03}

Because accurate (timing) positions are not yet available for many
of the radio pulsars, it is likely that some of them have been detected
in X-rays already but not yet identified as such. In fact, 
an X-ray source in NGC\,6397 was first identified with a possible 
BY Dra binary (Grindlay et al.\ 2001b); 
it was then found that this binary houses
a radio pulsar (Ferraro et al.\ 2001).
Similarly, NGC\,6752 CX11 was identified by Pooley et al.\ (2002a) with a
possible cataclysmic variable or background galaxy, but now is
more probably identified with PSR\,D in that cluster on the basis
of newly determined timing positions (D'Amico et al.\ 2002); 
positions of X-ray sources are coincident with the timing positions of PSRs C and 
(marginally) B.
\nocite{fpas01}\nocite{plh+02}\nocite{dpf+02}

Verbunt et al.\ (1996) showed that for the radio pulsars detected in
X-rays with ROSAT, $L_{0.1-2.4{\rm keV}}\ltap10^{-3}L_{sd}$, where
$L_{sd}\equiv I\Omega\dot\Omega$ is the loss of rotation energy,
usually referred to as the spin-down luminosity, with $I$ the moment
of inertia and $\Omega\equiv2\pi/P$.  In accordance with this scaling,
the radio pulsars detected in X-rays so far are those with the highest
$L_{sd}$ of those in the clusters observed with Chandra.  Grindlay et
al.\ (2002) assume that the electron density in 47\,Tuc is
homogeneous, and from small differences in dispersion measures
determine the position of each pulsar along the line of sight; this is
then used to correct the observed period derivative for gravitational
acceleration in the cluster potential. Comparison of the corrected
spindown luminosities with the X-ray luminosities led Grindlay et al.\
(2002) to suggest that $L_x \propto\sqrt{L_{sd}}$. They further note
that the pulsar in NGC\,6397 agrees with this (slower) trend, whereas
the pulsar in M\,28 does not. Grindlay et al.\ argue that the emission
of the pulsar in M\,28 is mainly magnetospheric in origin, whereas the
emission of the other pulsars in globular clusters is mainly thermal
emission from the surface of the neutron star.\nocite{vkb+96}\nocite{gch+02}

We reinvestigated the relation between X-ray and the spindown
luminosities for the globular cluster pulsars in
Figure\,\ref{verbflxe}. We include NGC\,6752\,D, and the pulsar in
M\,28. It should be noted that the luminosity of the pulsar in M\,28
is only about 20\%\ of the total cluster luminosity as observed with
ROSAT (Verbunt 2001; Becker et al.\ 2003), whereas the value used by
Grindlay et al.\ (2002) is the total cluster luminosity. Since thermal
emission from millisecond pulsars is the result of heating by
magnetospheric processes, we prefer not to exclude the magnetospheric
X-rays and to retain the pulsar in M\,28, and we are inclined to
conclude that the general slope of the relation between $L_x$ and
$L_{sd}$ is similar to that observed for the pulsars detected in the
Galactic disk, with some scatter at the lowest luminosities.
\nocite{ver01}\nocite{bsp+03}\nocite{gch+02}
The strong downward revision of the spindown luminosity of the
pulsar in NGC\,6397 (Possenti et al.\ quoted in Bassa \&\ Stappers
2004) brings this pulsar also in line with the steeper dependence of
$L_x$ on $L_{sd}$.  \nocite{vh98}

\subsection{Cataclysmic variables \label{seccv}}

\begin{figure}
\centerline{
\psfig{figure=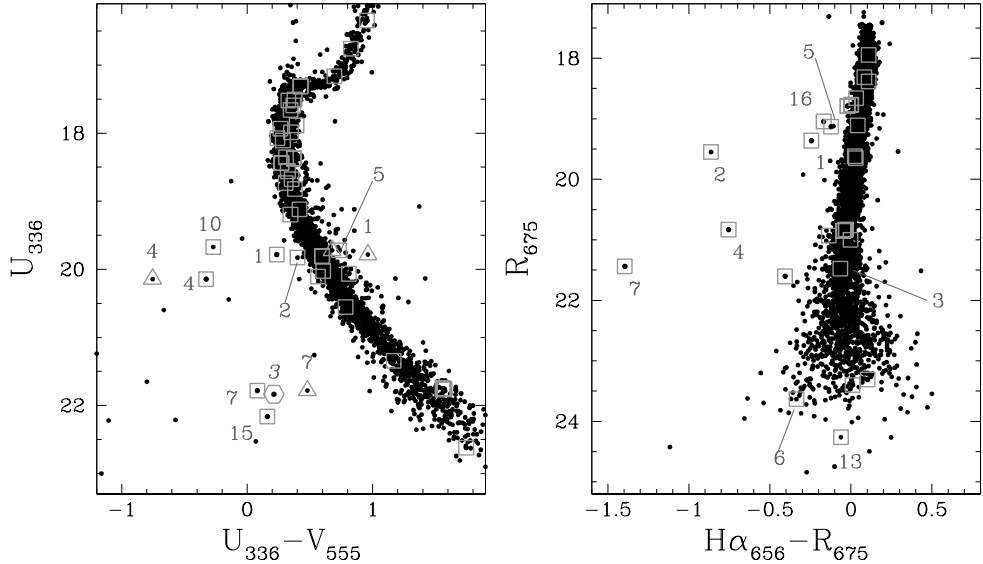,width=13.cm,clip=t}
}

\caption[o]{\it $U$--$V$ and $H\alpha$--$R$ color magnitude diagrams
of the central regions of NGC\,6752. Stars within error circles of
Chandra X-ray sources are indicated with squares; numbers indicate the
corresponding Chandra source. Cataclysmic variables lie to the left of
the main sequence in the $U$--$V$ diagram, i.e.\ they are blue.  When
the flux in the narrow $H\alpha$ filter is higher than in the
neighboring continuum (measured in $R$), the points fall to
the left
of the main sequence in the $H\alpha$--$R$ diagram.  Because of
variability, the same object may lie in different locations of the
color magnitude diagrams, depending on which data set is used. 
Updated after Pooley et al.\ (2002a).\label{verbfcmd}}
\end{figure}

Cataclysmic variables are best identified when an optical counterpart
is found. A good indicator is that the optical counterpart is bluer
than the main sequence, especially in the ultraviolet; and/or that it
has strong $H\alpha$ emission (see Figure\,\ref{verbfcmd}).  As an
example, such counterparts were identified in NGC\,6397, and follow-up
spectra show the strong Balmer emission lines prevalent in cataclysmic
variables (Cool et al.\ 1995, Grindlay et al.\ 1995, Edmonds et al.\
1999; note that firm identifications were only possible once Chandra
had obtained accurate positions, Grindlay et al.\ 2001b).  Quiescent
neutron-star low-mass X-ray binaries also have blue spectra with
Balmer emission, but can be distinguished from cataclysmic variables
through their soft X-ray spectra, and by the fact that they are more
luminuos than cataclysmic variables (see \S\ref{secflmx}).
Optical and ultraviolet color-magnitude diagrams have been used to
classify optical counterparts as cataclysmic variables also in
NGC\,6752 and in 47\,Tuc (Pooley et al.\ 2002a, Edmonds et al.\ 2003).
\nocite{cgc+95}\nocite{gcc+95}\nocite{egc+99}\nocite{ghem01}
\nocite{plh+02}\nocite{eghg03}

\nocite{plh+02}

\begin{figure}
\centerline{
\parbox[b]{8.0cm}{\psfig{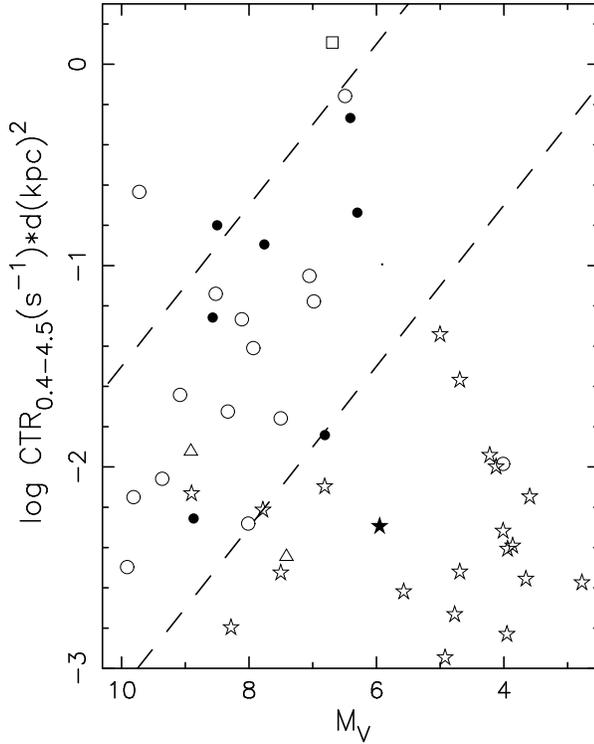}}
\parbox[b]{5.0cm}{\caption[o]{\it X-ray luminosity as a function of
absolute visual magnitude, for optically identified Chandra sources
in 47\,Tuc (open symbols) and NGC\,6752 (filled symbols). Squares, circles,
triangles and stars indicate low-luminosity \lmns, cataclysmic
variables, (companions to) recycled pulsars, and magnetically
active binaries, respectively. To minimize model dependence,
the X-ray luminosity is expressed as the product of Chandra countrate
CTR (in the 0.5--4.5 keV band, corrected for interstellar
absorption) and the cluster distance $d$ (in kpc)
squared. Two dashed lines of constant ratio of X-ray to visual flux
roughly separate the low-luminosity low-mass X-ray binaries with neutron
stars from the cataclysmic variables; and the latter from the 
magnetically active binaries (see Verbunt \&\ Hasinger 1998, Pooley
et al.\ 2002a). Data from Edmonds et al.\ (2003), Pooley et al.\ (2002a).
\label{verbfvlx}}
}
}
\end{figure}

If no optical colors are available, the ratio of X-ray to optical
flux provides a good, but not conclusive, indication as to whether
a source is a cataclysmic variable, as shown with cataclysmic
variables studied in the ROSAT All Sky Survey (Verbunt et al.\ 1997,
Verbunt \&\ Johnston 2000).
In Figure\,\ref{verbfvlx} we show (a measure of) the X-ray luminosity
in the 0.5--4.5\,keV range as a function of the absolute visual magnitude
for X-ray sources in 47\,Tuc and in NGC\,6752.
Only sources which have been classified on the basis of 
optical/ultraviolet color magnitude diagrams are shown.
In the figure we plot the line
\begin{equation} \log\left({\rm CTR}_{\rm 0.5-4.5keV} {d_{\rm kpc}}^2\right) =
- 0.4M_V + 0.9 \label{eqct}\end{equation}
where ${\rm CTR}_{\rm 0.5-4.5keV}$ is the number of counts per second
in the 0.5--4.5\,keV range, and $d_{\rm kpc}$ the distance in kpc.
This line roughly separates the cataclysmic variables from magnetically
active binaries. A parallel line for an X-ray luminosity which
is a factor $\simeq40$ higher roughly separates the cataclysmic
variables from the low-luminosity low-mass X-ray binaries with a neutron star.
The figure shows that the ratio of X-ray to optical luminosity is
a fairly good classifier of X-ray sources in the absence of more 
conclusive information.
\nocite{vbrp97}\nocite{vj00}

\nocite{eghg03}\nocite{plh+02}

A further indicator that a source is a cataclysmic variable may be
found from optical variability, either orbital or from a (dwarf) nova
outburst. Orbital variability may be present in magnetically active
binaries too, and thus can be used to classify a source only in
combination with other information, such as color magnitude diagrams,
or ratio of X-ray to visual flux. Two cataclysmic variables were found 
in NGC\,6752 based on periodic variability and $H\alpha$ emission 
by Bailyn et al.\ (1996), and were identified with Chandra X-ray sources
by Pooley et al.\ (2002a). Variability indicative of
dwarf nova outbursts has been detected for several blue objects
in 47\,Tuc (e.g.\ Paresce et al.\ 1992, Paresce \&\ De Marchi
1994, Shara et al.\ 1996); these sources have subsequently been 
identified with Chandra X-ray sources (Grindlay et al.\ 2001a).
An optical variable in the core of NGC\,6656/M\,22 has been identified as
a possible dwarf nova, detected in X-rays with Einstein, ROSAT
and XMM (Anderson et al.\ 2003; see Table\,\ref{tdim}).
\nocite{brs+96}\nocite{plh+02}\nocite{pm94}\nocite{pmf92}\nocite{sbg+96}
\nocite{ghem01}\nocite{ack03}

So far, only 47\,Tuc, NGC\,6397 and NGC\,6752 have been studied
to such an extent that a large fraction of the X-ray sources in
them has been optically identified. Most of them are classified as
cataclysmic variables. In $\omega$\,Cen, several Chandra sources
have been identified with (optically detected) cataclysmic variables
(Carson et al.\ 2000),
but HST observations only cover a small fraction of the cluster.
Classifications based only on the X-ray to optical flux ratio must
be considered preliminary, as illustrated by the case of NGC\,6752 CX11
(see \S\ref{secmsp}).
\nocite{ccg00}

In general it may be stated that the properties of cataclysmic variables
in globular clusters are similar to those of cataclysmic variables in
the Galactic disk (i.e.\ in the solar neighborhood; see also \S10\mk). In the
Galactic disk, distances and interstellar absorption for cataclysmic
variables are only inaccurately determined at best. In contrast,
for systems in globular clusters these quantities may be set equal
to the values for the cluster, which are much better known. Thus
comparison between different classes of objects will be more accurate
in globular clusters.

As an example, we note that Verbunt \&\ Hasinger (1998) in their
analysis of ROSAT observations of 47\,Tuc use the ratio of X-ray to
visual flux to suggest that 47\,Tuc X9, identified with the blue
variable V1, is a low-luminosity low-mass X-ray binary with a neutron
star. In Figure\,\ref{verbfvlx}, based on more accurate Chandra data
and now secure identifications, the systems with the three highest
X-ray to optical flux ratios in 47 Tuc are X10/V3, X7 and X9/V1. X7 is
indeed a low-luminosity low-mass X-ray binary with a neutron star, but
the hard X-ray spectra of X10 and X9 indicate that they are probably
cataclysmic variables. This illustrates the overlap between low-mass
X-ray binaries and cataclysmic variables in the X-ray to visual flux
ratio. 

\begin{table*}
 \begin{tabular}{l@{\hspace{0.2cm}}rrrl@{\hspace{0.2cm}}r@{\hspace{0.2cm}}rllrl}
cluster & ref & $L_{low}$ & BX & FX & CV & PSX & (PSR) & BY &
 $N_{tot}$ \\
NGC\,6440 & \cite{plv+02} & $2\times10^{31}$ & 1 & 3 & & & (1) \\
NGC\,6652 & \cite{heg01}  & $8\times10^{32}$  & 1 & 3 & $\rightarrow$1? \\
Terzan\,1 &  \cite{whg02}  & $3\times10^{33}$  & 1 & 1 & $\rightarrow$1? \\
Terzan\,5 &  \cite{heg+03} & $5\times10^{32}$  & 1 & 4 & 5 & & (4) \\
Liller\,1 &  \cite{hdam01} & $\sim10^{34}$  & 1 & 3 \\
\\
47\,Tuc  &  \cite{ghem01} & $10^{30}$ & 0 & 2 & $>$30 & 15 & (22) & 26 & 104 \\
$\omega$\,Cen &  \cite{gbw03} & $10^{31}$ & 0 & 1 & $>$20 & & (0) & 4 & $\sim$100 \\
NGC\,6093 &  \cite{hge+03} & $7\times10^{30}$ & 0 & 2 & $\sim$15 & & & & 19 \\
NGC\,6121 &  \cite{bph+04} & $1\times10^{29}$ & 0 & 0 & 3 & 1 & (1) & 14 & $\sim$20 \\
NGC\,6205 &  \cite{gbw03b} & $2\times10^{31}$ & 0 & 1 & 4 & 0 & (5) & & 5 \\
NGC\,6397 &  \cite{ghe+01} & $3\times10^{29}$ & 0 & 1 & 9 & 1 & (1) & 3 & $\sim$20 \\
NGC\,6626 &  \cite{bsp+03} & $2\times10^{30}$ & 0 & 1 & $\sim$25 & 1 & (1) & &  \\
NGC\,6656 &  \cite{wgb02} &          & 0 & 1?$\leftarrow$ & 3 & & & $\sim$3 \\
NGC\,6752 &  \cite{plh+02} & $2\times10^{30}$ &  0 & 0 & 10 & 1 & (5) & 3 & 17 \\
 \end{tabular}
 \caption[o]{\it Published Chandra and XMM observations of low-luminosity X-ray sources
in globular clusters. For each cluster we give the lowest detectable
luminosity (\ergsec, estimated for the range 0.5-2.5 keV), 
and the estimated numbers of X-ray sources corresponding to
 luminous low-mass X-ray binaries (BX),
low-luminosity low-mass X-ray binaries (FX), cataclysmic variables (CV),
recycled pulsars (PSX), (for comparison: we list the number of radio pulsars in the column PSR), and magnetically active binaries (BY). 
$\rightarrow$1? (1?$\leftarrow$) indicates that one of the sources
in the previous (next) column actually may belong in this column.
The final column gives the total number of detected X-ray sources
associated with the cluster.\label{verbta}}
 \end{table*}

\subsection{Magnetically active binaries}

X-ray sources in globular clusters can be classified as
magnetically active binaries when a stellar flare is observed in
X-rays; or on the basis of the optical counterpart, when this
is a known active binary, or less securely when it
lies above the main sequence and/or shows weak $H\alpha$ emission.

Two OGLE variables in NGC\,5139, OGLEGC15 and OGLEGC22, are identified
by Cool et al.\ (2002) with Chandra sources (not listed by
Rutledge et al.\ 2002, but confirmed by
Gendre, private communication). A third OGLE variable in NGC\,5139, OGLEGC30,
has been \mk detected with XMM (Gendre et al.\ 2003a). 
Yet another Chandra X-ray source, already detected
with ROSAT but not detected with XMM and therefore a variable X-ray source,
shows $H\alpha$ emission, and presumably is also a magnetically active 
binary (Gendre et al.\ 2003a).
Figure\,\ref{verbfvlx} shows Chandra X-ray sources in 47\,Tuc and NGC\,6752
that are classified on the basis of color-magnitude diagrams
as magnetically active binaries; for many of these binaries in 47 Tuc
the orbital lightcurve confirms their identity as coronal X-ray
emitters (Edmonds et al.\ 2003).
That care must be taken in classifying sources is shown by the example
of NGC\,6397 CX12 (see \S\ref{secmsp}).
\nocite{chc02}\nocite{rbb+02}\nocite{gbw03}

Interestingly, most magnetically active binaries identified 
with X-ray sources so far have visual magnitudes higher than or 
equal to the turnoff stars, implying that they are on the main sequence
(BY\,Dra's). Since the maximum X-ray luminosity
of a magnetically active binary scales roughly with the surface
area of the stars, this implies that the luminosities of the active
binaries in globular clusters are low
(typically $L_x<10^{30}$\ergsec), compared to systems with giants
(RS CVn's), in the
Galaxy, which can be up to a hundred times more luminous (Dempsey et 
al.\ 1993).
\nocite{dlfs93}

\subsection{Comparing clusters}

In comparing the different clusters, the limit to which sources can be
detected must be taken into account. Low-luminosity low-mass X-ray
binaries with a neutron star tend to be more luminous than cataclysmic
variables, which in turn tend to be more luminous than magnetically
active binaries.  This ordering is reflected in the numbers of
currently known cataclysmic variables and magnetically active binaries
listed in Table\,\ref{verbta} as a function of the detection limit.

\begin{figure}[ht!]
\begin{center}
\resizebox{1.0\textwidth}{!}{\includegraphics{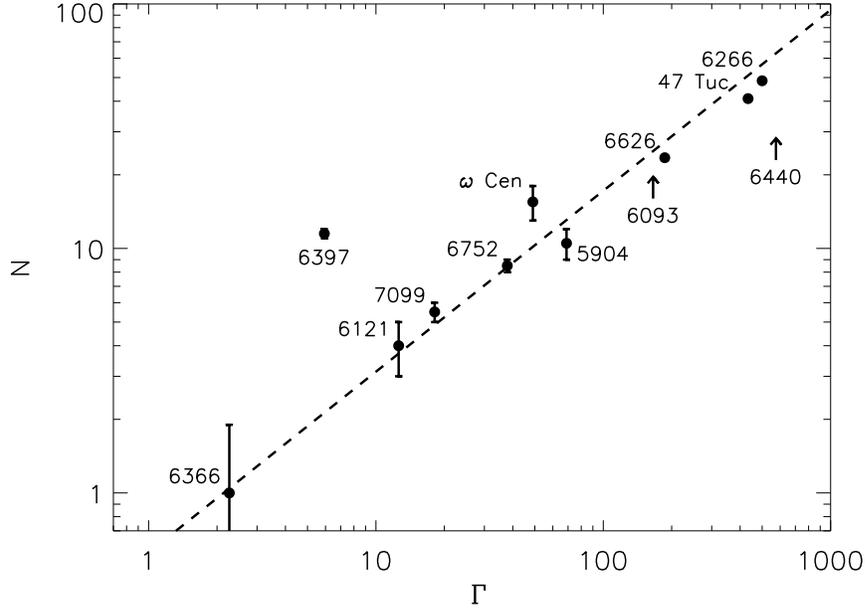}}
{
\caption{\label{collfreq} Number $N$ of X-ray sources with
$L_x\gtap4\times10^{30}$~\ergsec\ (0.5-6 keV) detected in globular clusters,
as a function of the collision number $\Gamma$. $\Gamma$ is a measure
of the number of close encounters between stars in a cluster
(see Eqs.\,\ref{gamma} and \ref{gammae}).
The luminosity limit implies that most sources are cataclysmic variables.
In general $N$ scales quite well with $\Gamma$, indicating that 
cataclysmic variables in globular clusters are formed via close
encounters between a white dwarf and another star or a binary.
Arrows indicate lower limits. NGC\,6397 doesn't follow the general
trend. From Pooley et al.\ (2003).\label{Pooley}}}
\end{center}
\end{figure}

Another number that is important is the estimated number of
close encounters between stars in the globular cluster. Pooley et 
al.\ (2003) show that the number of X-ray sources detected in a 
globular cluster above an observational threshold of
$L_x\simeq4\times10^{30}$~\ergsec\ (0.5-6 keV) scales quite well with 
this number, as shown in Figure\,\ref{Pooley}. Heinke et al.\ (2003d)
find that the number of cataclysmic variables alone  
(at $L_x\gtap\times10^{31}$~\ergsec) possibly increases slower
with central density than predicted by proportionality to the
number of close encounters.
\nocite{pla+03}\nocite{hgl+03}

An exception to this scaling is NGC\,6397. This cluster has a 
higher number of neutron star binaries and cataclysmic variables
than expected on the basis of its rather low collision number.
Remarkably, the number of magnetically active binaries in this cluster is not
very high, and this is reflected in a relatively flat X-ray
luminosity function (Pooley et al.\ 2002b). If it is true, as
argued by Pooley et al.\ (2003), that the high number of
neutron star binaries and cataclysmic variables
in NGC\,6397 is due to its being shocked and 
stripped in multiple passages through the galactic disk and/or near the galactic center, it has to
be explained why these mechanisms are more efficient in removing
magnetically active binaries than in removing cataclysmic variables
and binaries with neutron stars.

\section{Some remarks on evolution and formation}

\subsection{Evolution}

\begin{figure}
\centerline{\psfig{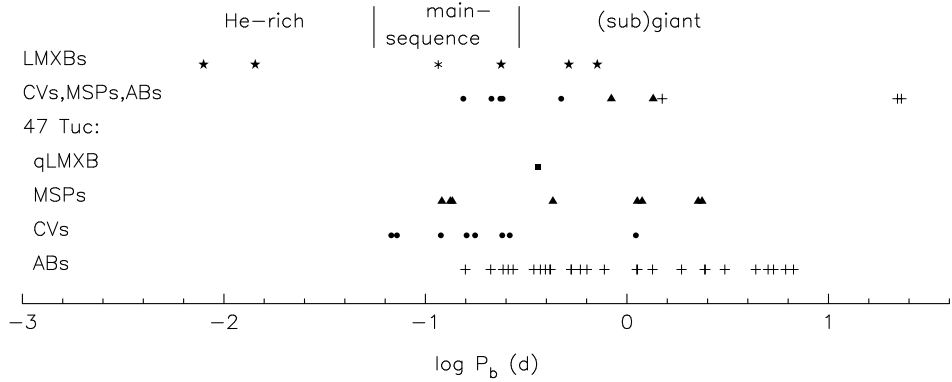}}

\caption[o]{\it Orbital period distributions 
of X-ray-detected binaries in globular 
clusters. Most known orbital periods are for systems in 47\,Tuc, and are
shown in the lower four rows. The top two rows
indicate the luminous X-ray binaries and other binaries 
in other clusters (with symbols as for 47\,Tuc). The period of
a cluster source in M\,31 is shown with a six-pointed star.
The period range in which a main-sequence star can fill its
Roche lobe is indicated; systems with shorter periods
may contain degenerate stars, with longer periods
(sub)giants. Periods from Table\,\ref{tbright}; 47\,Tuc: Edmonds et 
al.\ (2003), Freire et al.\ (2003), Camilo et al.\ (2000); other
clusters: Bailyn et al.\ (1996),  Deutsch et al.\ (2000), Neill et al.\ (2002), Kaluzny \&\ 
Thompson (2002), Kaluzny et al.\ (1996), D'Amico et al.\ (2001, 2002); 
M\,31: Trudolyubov et al.\ (2002).
\label{verbfper}}
\end{figure}

A good first indicator of the evolutionary status of a binary is its
orbital period (see Chapter 16 by Tauris \&\ Van den Heuvel and
Verbunt 1993 for a more extended discussion of the evolution of X-ray
binaries).  We show the orbital periods of X-ray emitting binaries in
globular clusters in Figure\,\ref{verbfper}. Most periods known are
for binaries in 47\,Tuc. It should be noted that there is a selection
effect against the discovery of long-period binaries in optical
surveys.

The radius $R$ of the Roche lobe of a star with mass $M$ in a binary 
with a star of mass $m$, is given in units of the distance $a$ between
stars as approximately 
\begin{equation}
{R\over a} \simeq 0.46 \left({M\over M+m}\right)^{1/3} \qquad
{\rm for}~M<0.8m
\end{equation}
Combining this with the third law of Kepler we find
\begin{equation}
P_b \simeq 8.9\,{\rm hr} \left({\Msun\over M}\right)^{1/2}
\left({R\over\rsun}\right)^{3/2}
\end{equation}
i.e.\ the orbital period gives the average density of a Roche-lobe
filling star (cf. \S5.3.1)\mk.
\nocite{ver93}

\nocite{eghg03}\nocite{fck+03}\nocite{clf+00}\nocite{brs+96}\nocite{nscb02}
\nocite{dpm+01}\nocite{dpf+02}\nocite{tbp+02}

The radius of a main sequence star is roughly given by 
$R/\rsun\simeq M/\Msun$ in the mass range of interest here.
With main-sequence stars in old globular clusters limited to masses
$M\ltap0.8\rsun$, we see that binaries in which mass transfer occurs,
i.e.\ low-mass X-ray binaries and cataclysmic variables, can only
have a main sequence star as the mass donor provided the orbital period
is less than about 7~hr. If the orbital period is longer, the donor
must be larger than a main-sequence star, i.e.\ a (sub)giant.
It then follows from Figure\,\ref{verbfper} that, with one exception,
all cataclysmic variables in globular clusters can have main-sequence
donors. The one exception is AKO9, a cataclysmic variable with a
slightly evolved donor in 47\,Tuc (e.g.\ Knigge et al.\ 2003). \nocite{kzs+03}
Of the low-mass X-ray binaries, one may have a main-sequence donor,
two binaries must have subgiant donors; the low-luminosity low-mass X-ray binary
in 47\,Tuc is probably a subgiant close to the main sequence.
The orbital periods of most active binaries are long enough that
even main-sequence stars near the turnoff mass ($0.8\Msun$) fit
well within the Roche lobes; for those with the shorter periods,
both stars must have lower masses to be smaller than their 
Roche lobes.
Two of the low-mass X-ray binaries have ultra-short orbital periods;
at such short orbital periods the Roche filling star can be a white dwarf.
With $R/\rsun\simeq0.01(M/\Msun)^{-1/3}$, a white dwarf fills its
Roche lobe if the orbital period $P_b\simeq48\,{\rm s}\Msun/M$.

The evolution of low-mass X-ray binaries and cataclysmic variables
{\em with main-sequence donors} is driven by the loss of angular
momentum $\dot J$ from the binary. Writing the angular momentum of the
binary as $J_b$, one finds that the mass-transfer rate $\dot M$ is
roughly given by $-\dot M/M\sim-\dot J/J_b$. The loss of angular
momentum from gravitational radiation alone is enough to drive mass
transfer at a rate of $10^{-10}\Msun yr^{-1}$; higher mass transfer
rates, as witnessed by luminosities well in excess of
$L_x\simeq10^{36}$~\ergsec, imply other mechanisms. The loss of
angular momentum causes the orbit to shrink, and thus the orbital
period to become shorter.  In binaries {\em with a (sub)giant donor},
the mass transfer rate is very roughly given by the expansion rate of
the donor star $-\dot M/M\sim\dot R/R$. Since the expansion rate of a
giant becomes faster as it further ascends the giant branch, this
predicts higher mass transfer, i.e.\ more luminous X-ray emission, for
the longest periods. For the two known orbital periods of low-mass
X-ray binaries in globular clusters with a subgiant, expansion of the
donor predicts a modest mass transfer of $\sim10^{-10}\Msun yr^{-1}$.
The mass transfer, combined with conservation of angular momentum,
causes the orbit to expand, and the orbital period to increase.
Enhanced loss of angular momentum from a stellar wind has often been
invoked to explain large X-ray luminosities, in binaries with
main-sequence or subgiant donors, but the actual efficiency of this
loss mechanism is not known.  It is worth noting that many X-ray
sources show large variations in their X-ray luminosity on time scales
of decades -- the transients are an obvious example -- indicating that
the current mass transfer rate, even in apparently stable systems, may
not be an accurate estimator of mass transfer rate on an evolutionary
time scale.

Something is wrong with the simplest description of binary
evolution. This follows, e.g., from the orbital period distribution
of the recycled radio pulsars. The expansion of a binary with a subgiant 
donor continues until the core of the giant is denuded of its envelope.
By then the orbital period has increased by an order of magnitude.
The orbital periods of the radio pulsars in 47 \,Tuc are less than
about 2.5\,d, suggesting that little if any expansion has occurred 
during the mass transfer. On the other hand, some pulsar binaries
in globular clusters, such as the pulsar binary in M\,4, do have 
periods in excess of hundred days,
with fairly circular orbits, showing that expansion is strong in at
least some cases. 

What about the ultrashort periods? They may have white-dwarf donors;
if so, their orbital period should be increasing.
It has been suggested that a collision between a (sub)giant and a
neutron star could lead to expulsion of the giant envelope and leave
the neutron star in orbit around the core, which subsequently cools
to an under-massive white dwarf. If loss of angular momentum
from gravitational radiation pushes the stars closer, mass transfer
begins once the white dwarf fills its Roche lobe (Verbunt 1987).
Alternatively, it has been suggested that the ultrashort
period systems are the outcome of an evolution which starts when
a subgiant starts transferring mass to a neutron star in an
orbital period less than $\sim18$\,hr (Podsiadlowski et al.\ 2002). 
Large loss of angular momentum
through a stellar wind brings the two stars closer together,
and the evolution proceeds to shorter and shorter periods. The minimum
period reached through such an evolutionary path is short enough to
explain the 11\,min period of the \lmns\ in NGC\,6624.
It is predicted that this binary has a
negative period derivative, as observed. 
There are two problems with this scenario, however. One is that the
loss of angular momentum from the giant, required at the start of the
mass transfer to convert orbital expansion into orbital shrinking,
is rather high; perhaps implausibly high. Second, none of the
evolutions along this scenario computed by Pylyser \&\ Savonije (1988) 
reach the shortest periods within a Hubble time, because it already takes 
very long for a 1$\Msun$ star to fill its Roche lobe in a 16\,hr period.
Van der Sluys et al.\ (2005) investigate this in more detail and find
that only binaries in narrow ranges of initial orbital periods and component
masses evolve to ultrashort periods within a Hubble time, and that these
binaries only spend a small fraction of their life at ultrashort periods;
they conclude that no significant population of ultrashort-period binaries 
in globular clusters can be produced through this evolution channel.
\nocite{ver87}\nocite{ps88}\nocite{prp02}
The most likely mechanism to produce bright X-ray sources with ultrashort
orbital periods is mass transfer from an intermediate-mass donor leading
to a common envelope, some time in the past history of the globular
cluster (Davies \&\ Hansen 1998, Rasio et al. 2000). The result is a
binary of the neutron star and the core of the
giant, which cools into a white dwarf. In the course of several billion
years, gravitational radiation may bring the system into contact.

\subsubsection{Some specific systems}

The orbital period for the low-luminosity low-mass X-ray binary
47~Tuc X5 is too long for a Roche-lobe filling
main sequence donor star with a mass less than the turnoff mass of
$0.8\Msun$.
Edmonds et al.\ (2002b) therefore conclude that the star is smaller
than its Roche lobe. We suggest an alternative possibility that the system
hosts
a 0.8~$\Msun$ subgiant donor
that has recently started to transfer matter to a 1.4~$\Msun$ neutron
star.
The donor has not yet transferred much of its envelope mass:
a low donor mass in an 8.666 hr orbit implies a Roche lobe for the
donor that is too small to hold a subgiant. The system is very sub-luminous
for a subgiant: this is expected for a donor that is losing mass.
\nocite{ehgg02}

PSR 47\,Tuc\,W (Chandra source 29) is a pulsar accompanied by an
object whose location in the color-magnitude diagram indicates that
it is too big for a white dwarf and too small for a main-sequence
star. The orbital lightcurve shows clear heating by the pulsar
(Edmonds et al.\ 2002a).  If a main-sequence star is heated at constant
radius, it moves up and to the left in a color-magnitude diagram, to
a location below the main-sequence.  If the companion to PSR
47\,Tuc\,W is of this nature, its position about 5 magnitudes below
turnoff indicates a very low mass, of an M dwarf.  This poses an
interesting puzzle for the evolutionary history: if the M dwarf was in
the binary from the start, it was too small to transfer mass to the
neutron star and spin it up. If on the other hand the main-sequence
star was captured by the pulsar tidally or via an exchange encounter,
the orbit should be eccentric initially; the question is whether
tidal dissipation can circularize the orbit and heat the M dwarf to
its current position.
\nocite{egc+02}

PSR NGC\,6397\,A is another pulsar accompanied by a low-mass
($\sim 0.25\Msun$) companion (Ferraro et al.\ 2003). In this case
the companion lies somewhat to the right of the turnoff, at a
radius of 1.6(2)$\rsun$ and luminosity 2.0(4)$\lsun$; notwithstanding
the proximity of an energetic radio pulsar, the companion shows no
sign of heating (Orosz \&\ van Kerkwijk 2003). The position
of the companion in the color-magnitude diagram is hard to explain.
Orosz \&\ van Kerkwijk invoke a stellar collision, causing a slightly
evolved star near the turnoff to lose most of its envelope.
\nocite{fsg+03}\nocite{ok03}

\subsubsection{Black holes\label{vlsecbh}}

The absence of known very luminous ($L_x\geq10^{38.5}$~\ergsec,say)
low-mass X-ray binaries with a black
hole in globular clusters of our Galaxy has led to the suggestion
that black holes are efficiently ejected from globular clusters
through dynamical processes (Kulkarni et al.\ 1993; Portegies Zwart
\&\ McMillan 2000). The discovery of very luminous, soft X-ray
sources in globular clusters in other galaxies shows that X-ray
binaries with black holes probably exist in globular clusters (see \S8.3).
\nocite{khm93}\nocite{pm00}

There is no evidence that M\,15
contains an intermediate mass black hole;
an upper limit for the mass
of about $10^3\Msun$ can be set both from an  analysis of pulsar
accelerations in this cluster, and from an analysis of radial velocities
of stars close to the center (Phinney 1992; Gerssen et al.\ 2003).
A case has been made for a binary in  NGC\,6752 of two black holes, of which 
at least one has an intermediate mass (Colpi et al.\ 2002).
The argument for this is the presence of a white-dwarf/radio-pulsar binary in the
outskirts of the cluster, which most likely was ejected from the
cluster core. If the binary was ejected 
with the white dwarf companion to the pulsar already formed,
the very small eccentricity of its orbit implies that the orbit
of the other binary involved in the scattering was much larger.
To still produce an ejection velocity for the pulsar binary
high enough for it to reach the outer cluster region then requires
at least one black hole with a mass $\sim100\Msun$ in the
scattering binary (Colpi et al.\ 2002).
To solidify the case for a binary black hole it would have to
be demonstrated that the pulsar indeed belongs to NGC\,6752 (as is
probable), and that the pulsar binary was ejected before the
formation of the white dwarf (which is not obvious).
The optical identification of the white dwarf companion to this
pulsar shows that the white dwarf is young compared to the
age of the globular cluster; this strengthens the case for a
scenario in which a binary consisting of a main-sequence star 
and a neutron star was ejected from the cluster core, and
subsequent evolution of the main-sequence star led to circularization
of the orbit (Bassa et al.\ 2003).
\nocite{phi92}\nocite{gmg+03}\nocite{cpg02}\nocite{bvkh03}

\subsection{Formation}

The rate at which stars with number density $n$ encounter target stars
with number density $n_c$ in a cluster with dispersion velocity $v$
is given by (e.g.\ Hut \&\ Verbunt 1983):
\begin{equation}
\Gamma \propto \int n_cnAv dV \propto \int {n_cnR\over v} dV
 \propto {\rho_o^2r_c^3\over v}\,R
\label{gamma}
\end{equation}
where $A$ is the interaction cross section (proportional to $R/v^2$ because
of gravitational focusing), $R$ the radius of the star,
$\rho_o$ is the central mass density \mk of the cluster and $r_c$ 
its core radius. Because the number densities of stars drop rapidly
with distance from the cluster center, the integral over volume $dV$
can be approximated by multiplying the central encounter rate with the
volume of the cluster core.
\nocite{hv83}
An analogous equation gives the exchange encounter rate
\begin{equation}
\Gamma_e \propto \int n_cn_bA_bv dV \propto \int {n_cn_ba\over v} dV
 \propto {\rho_o^2r_c^3\over v}\,a
\label{gammae}
\end{equation}
where $n_b$ is the number of binaries per unit volume, and $a$
the semi-major axis of the binary.
The ratio of tidal capture to exchange encounters is roughly
\begin{equation}
{\Gamma\over\Gamma_e} \sim {R\over a} {n\over n_b}
\end{equation}

The velocity dispersion $v$ is related to the core mass and radius
through (a specific version of) the virial theorem (King 1966):
\begin{equation}
v\propto \sqrt{\rho_o}\,r_c
\label{vel}\end{equation}
Therefore (Verbunt 2003)
\begin{equation}
\Gamma \propto {\rho_o}^{1.5}{r_c}^2R \qquad {\rm and} \qquad
\Gamma_e \propto {\rho_o}^{1.5}{r_c}^2a
\label{gammc}\end{equation}
\nocite{kin66}

Because neutron stars are formed with appreciable velocities, a
cluster with a high mass is expected to retain a higher fraction of
the neutron stars that are formed in it than a cluster with a low
mass. In a cluster with strong mass segregation, virtually all the
neutron stars will have migrated to the core. Thus a massive cluster
with strong mass segregation is expected to have a much higher central
number density of neutron stars than a low-mass little-segregated
cluster.  Thus, the ratio $n_c/\rho_o$ for neutron stars, and through
this the proportionality constant for the last members of
Eqs.\,\ref{gamma} and \ref{gammae} will vary widely between clusters
(Verbunt \&\ Meylan 1988). On the other hand, white dwarfs are always
retained upon formation, and due to their lower masses are less
affected by mass segregation. This is probably the reason that the
relation between the number of X-ray sources (mainly cataclysmic
variables) and $\Gamma$ is as narrow as shown in Figure\,\ref{Pooley}.
\nocite{vm88}

Due to the large number density of stars in a cluster core, an
appreciable fraction of neutron stars in that core may be involved in
a close encounter with a single star or with a binary. The formation
of tidal bulges during passage of a neutron star within $\sim3$ times
the radius of a main-sequence star drains enough energy from the
relative motion of the two stars to bind them in a binary. This process is
called tidal capture (Fabian et al.\ 1975). Whether it is efficient in the
formation of a binary with a neutron star is under debate, because
of the large amount of energy residing in the initially very eccentric
orbit of the newly formed binary. If the orbit circularizes rapidly
because of tides on the main sequence star, the energy released is
enough to (almost) destroy the main sequence star (Ray et al.\ 1987,
Verbunt 1994). Rapid circularization
can be avoided if the energy exchange between tides and orbit is
chaotic, as is likely in a highly eccentric orbit (Mardling 1995).
Mass loss from the main-sequence star due to tidal heating may further
limit the damage to the deeper layers of the star.
\nocite{rka87}\nocite{ver94a}\nocite{mar95}

A neutron star can also be exchanged into a pre-existing binary when
it takes the place of one of the binary members in an exchange
encounter (Hills 1976). Which of the two mechanisms is more important
depends on the number of binaries present in the core and on their
orbital period distribution; as well as on the efficiency of the tidal
capture process.
\nocite{hil76}

If a binary is of a type that very rarely results from the 
evolution of a primordial
binary, then its presence in a globular cluster may be ascribed to
formation via a close encounter. Such is the case for binaries
with a neutron star.
If a binary is very frequently formed from a primordial binary, then it
is likely to be primordial also when present in a globular cluster.
This is the case for magnetically active close binaries.
Cataclysmic variables are somewhere in between, and thus in clusters
can be formed both via close encounters and via evolution of a primordial
binary. Figure\,\ref{Pooley} shows that the number of binaries with
$L_x\gtap4\times10^{30}$~\ergsec\ scales well with the number of encounters
in a cluster. Since most binaries with such luminosities are cataclysmic
variables, this implies that most cataclysmic variables are in fact formed
via close encounters. One reason for this is that evolution from a primordial
binary into a cataclysmic variable passes through a stage in which the
binary is very wide; such a wide binary is easily unbound in a globular
cluster by a passing star and the formation of a cataclysmic variable
is prevented (Davies 1997). If the number of cataclysmic variables
increases more slowly with central density than as $\rho_o^{1.5}$,
as suggested by Heinke et al.\ (2003d), this could suggest that primordial
binaries do still contribute to the formation of cataclysmic variables.
Remarkably, Jord\'an et al.\ (2004) find that the probability for a
globular cluster associated with NGC\,4486 (M\,87) to harbor a bright
X-ray source also scales with a lower power of $\rho_o$ than the 
collision number, i.e.\ roughly as $\propto\Gamma\rho_o^{-0.5}$. 
\nocite{dav97}

Looking now at the period distribution of the cataclysmic variables
and low-mass X-ray binaries in globular clusters, we see that their
periods are short, $\ltap1$\,d. This may indicate that they are
formed at short periods, which hints at tidal capture as the main
formation process. Some care is necessary before one jumps to conclusions,
however. Mass transfer in wide binaries tends to be faster, and thus
wide binaries live shorter, and will be less numerous even if their
formation rate is the same as that of short binaries. Also, longer 
periods are more difficult to measure, and some of the many binaries
with unknown periods may have long periods.
In addition, a wide binary with a neutron star or white dwarf can
become closer via encounters with field stars. We doubt that this process
is sufficiently efficient, given the observed presence in 47\,Tuc 
of active binaries with periods up to ten days that apparently
have avoided further shrinking of their orbits.
On the whole we tend to conclude that rumors of the death of the tidal 
capture model for the formation of binaries with a neutron star and
of cataclysmic variables have been much exaggerated.

That exchange encounters do occur in globular clusters is evident from
the wide pulsar binaries, such as M\,4 PSR\,A ($P_b=191$\,d, Thorsett
et al.\ 1999) and M\,53 PSR\,A (255\,d, Kulkarni et al.\ 1991). 
These are found in clusters with a relatively
low central density, which allows long period binaries to survive
(e.g.\ Verbunt 2003). They must have evolved from binaries with
initial periods too long to be formed by tidal capture, in which the
neutron star can thus only have entered via an exchange encounter.
(M\,15 PSR\,C is an eccentric binary of two neutron stars in the outskirts
of M\,15, and is another product of an exchange encounter:
Phinney \&\ Sigurdsson 1991.) 
\nocite{tacl99}\nocite{kapw91}\nocite{ver03}\nocite{ps91}

\section*{Acknowledgements}
We are very grateful for comments, suggestions and help from Lorella
Angelini, Keith Ashman, Pauline Barmby, Cees Bassa, Boris Dirsch,
Rosanne Di Stefano, Josh Grindlay, Bill Harris, Piet Hut, Andres
Jord\'an, Arunav Kundu, Erik Kuulkers, Tom Maccarone, Dave Pooley,
Katherine Rhode, Craig Sarazin, Rudy Wijnands, and Steve Zepf.

{}

\end{document}